\documentclass[onecollarge,fleqn,runningheads, preprint,showpacs,preprintnumbers,amsmath,amssymb]{svjour2}
\smartqed  
\usepackage{graphicx}
\usepackage{dcolumn}
\usepackage{bm}
\journalname{Theoretical and Computational Fluid Dynamics}
\begin{document}

\title{Lyapunov Analysis for Fully Developed Homogeneous Isotropic Turbulence
}


\author{Nicola de Divitiis
}


\institute{Department of Mechanics and Aeronautics\\
University "La Sapienza", Rome, Italy \at
           via Eudossiana, 18  \\
              Tel.: +39-06-44585268\\
              Fax: +39-06-4881759\\
              \email{dedivitiis@dma.dma.uniroma1.it}           
}

\date{Received: date / Accepted: date}

\maketitle

\newcommand{\no}{\noindent}
\newcommand{\be}{\begin{equation}}
\newcommand{\ee}{\end{equation}}
\newcommand{\bea}{\begin{eqnarray}}
\newcommand{\eea}{\end{eqnarray}}
\newcommand{\bc}{\begin{center}}
\newcommand{\ec}{\end{center}}

\newcommand{\calr}{{\cal R}}
\newcommand{\calv}{{\cal V}}

\newcommand{\bff}{\mbox{\boldmath $f$}}
\newcommand{\bfg}{\mbox{\boldmath $g$}}
\newcommand{\bfh}{\mbox{\boldmath $h$}}
\newcommand{\bfi}{\mbox{\boldmath $i$}}
\newcommand{\bfm}{\mbox{\boldmath $m$}}
\newcommand{\bfp}{\mbox{\boldmath $p$}}
\newcommand{\bfr}{\mbox{\boldmath $r$}}
\newcommand{\bfu}{\mbox{\boldmath $u$}}
\newcommand{\bfv}{\mbox{\boldmath $v$}}
\newcommand{\bfx}{\mbox{\boldmath $x$}}
\newcommand{\bfy}{\mbox{\boldmath $y$}}
\newcommand{\bfw}{\mbox{\boldmath $w$}}
\newcommand{\bfk}{\mbox{\boldmath $\kappa$}}

\newcommand{\bfA}{\mbox{\boldmath $A$}}
\newcommand{\bfD}{\mbox{\boldmath $D$}}
\newcommand{\bfI}{\mbox{\boldmath $I$}}
\newcommand{\bfL}{\mbox{\boldmath $L$}}
\newcommand{\bfM}{\mbox{\boldmath $M$}}
\newcommand{\bfS}{\mbox{\boldmath $S$}}
\newcommand{\bfT}{\mbox{\boldmath $T$}}
\newcommand{\bfU}{\mbox{\boldmath $U$}}
\newcommand{\bfX}{\mbox{\boldmath $X$}}
\newcommand{\bfY}{\mbox{\boldmath $Y$}}
\newcommand{\bfK}{\mbox{\boldmath $K$}}

\newcommand{\bfrho}{\mbox{\boldmath $\rho$}}
\newcommand{\bfchi}{\mbox{\boldmath $\chi$}}
\newcommand{\bfphi}{\mbox{\boldmath $\phi$}}
\newcommand{\bfPhi}{\mbox{\boldmath $\Phi$}}
\newcommand{\bflambda}{\mbox{\boldmath $\lambda$}}
\newcommand{\bfxi}{\mbox{\boldmath $\xi$}}
\newcommand{\bfLambda}{\mbox{\boldmath $\Lambda$}}
\newcommand{\bfPsi}{\mbox{\boldmath $\Psi$}}
\newcommand{\bfomega}{\mbox{\boldmath $\omega$}}
\newcommand{\bfeps}{\mbox{\boldmath $\varepsilon$}}
\newcommand{\bfepsn}{\mbox{\boldmath $\epsilon$}}
\newcommand{\bfzeta}{\mbox{\boldmath $\zeta$}}
\newcommand{\bfkappa}{\mbox{\boldmath $\kappa$}}
\newcommand{\itPsi}{\mbox{\it $\Psi$}}
\newcommand{\itPhi}{\mbox{\it $\Phi$}}
\newcommand{\bint}{\mbox{ \int{a}{b}} }
\newcommand{\ds}{\displaystyle}
\newcommand{\Sum}{\Large \sum}

\begin{abstract}
The present work studies the isotropic and homogeneous turbulence for incompressible fluids through a specific Lyapunov analysis, assuming that the turbulence is due to the bifurcations associated to the Navier-Stokes equations.

The analysis consists in the 
calculation of the velocity fluctuation through the Lyapunov analysis of the local deformation
and the Navier-Stokes equations and in the study of the mechanism of the energy cascade
from large to small scales through the finite scale Lyapunov analysis of the relative motion between two particles.

The analysis provides an explanation for the mechanism of the energy cascade,
leads to the closure of the von K\'arm\'an-Howarth equation, and describes the 
statistics of the velocity difference.

Several tests and numerical results are presented.

\keywords{Bifurcations \and Lyapunov Analysis \and von K\'arm\'an-Howarth equation \and Velocity difference statistics}
\PACS{47.27.-i}
\end{abstract}

\section{Introduction \label{s1}}


In this work a novel procedure based on a specific Lyapunov analysis is presented
for studying the incompressible isotropic and homogeneous turbulence in an infinite domain.
The analysis is mainly  motivated by the fact that in turbulence the kinematics 
of the fluid deformation is subjected to bifurcations \cite{Landau44} and exhibits a chaotic behavior and huge mixing \cite{Ottino89}, \cite{Ottino90}, resulting to be much more rapid than the fluid 
state variables.
This characteristics implies that the accepted kinematical hypotheses for deriving the Navier-Stokes equations could require the consideration of very small length scales and times for describing the fluid motion \cite{Truesdell77} and therefore a very large number of degrees of freedom. 
\\
As well known, other peculiar characteristics of the turbulence are the mechanism of the kinetic energy cascade, directly related to the relative motion of a pair of fluid particles \cite{Richardson26}, \cite{Kolmogorov41}, \cite{Karman38}, \cite{Batchelor53} and responsible for the shape of 
the developed energy spectrum, and the non-gaussian statistics of the velocity difference.

This energy spectrum can be calculated through a proper closure of the von K\'arm\'an-Howarth equation  (see Appendix) or of its Fourier Transform \cite{Karman38}, \cite{Batchelor53}.
The equation describes the evolution of the correlation function $f$ of the longitudinal velocity $u_r$, and depends upon the term $K$ (see Appendix), directly related to the longitudinal triple-velocity correlation $k$.
This latter, due to the inertia forces, does not change the kinetic energy of the fluid and satisfies the detailed conservation of energy \cite{Batchelor53} which states that the exchange of energy between wave-numbers is only related to the amplitudes of these wave-numbers and of their difference \cite{Onsager06}.

Various authors (see for instance  \cite{Hasselmann58}, \cite{Millionshtchikov69}, \cite{Oberlack93}), propose, for $k$, the following diffusion approximation 
\bea
\ds k = 2 \frac{D}{u} \frac{\partial f} {\partial r}
\label{diffus}
\eea
where $r$ and $D = D(r)$ are the separation distance and the turbulent diffusivity,
whereas $u^2 =\langle u_i u_i \rangle/3 $ represents the longitudinal velocity standard deviation.
This implies that the closed von K\'arm\'an-Howarth
 equation is a parabolic equation in any case (also for $\nu$ =0).

To the author knowledge, Hasselmann in 1958 \cite{Hasselmann58} was the first that proposed a link between $k$ and $f$, using a simple model which expresses $k$ in function of the momentum convected through the surface of a spherical volume.
His model incorporates a free parameter and expresses $D(r)$ by means of a complex expression.

Another closure model in the framework of Eq. (\ref{diffus}) 
was developed by Millionshtchikov \cite{Millionshtchikov69}.
There, the author assumes that  
$
\ds D(r) = k_1 u \ r
\label{Millionshtchikov}
$,
where $k_1$ is an empirical constant. 
Although both the models describe two possible mechanisms of the energy cascade, in general, do not
satisfy some physical conditions. For instance, the model of Hasselmann does not verify the
continuity equation for all the initial conditions, whereas the Millionshtchikov's model gives, according to Eq. (\ref{diffus}), an absolute value of the skewness of $\partial u_r/ \partial r$, in contrast with the several experiments and with the energy cascade \cite{Batchelor53}.

More recently, Oberlack and Peters \cite{Oberlack93} suggested a closure model where $D$ is in terms of $f$, i.e. 
$
D(r)  = k_2 r \ u \sqrt{1- f} 
\label{Oberlack}
$  
and $k_2$ is a constant parameter.
The authors show that this closure reproduces the energy 
cascade and, for a proper choice of $k_2$,  provides results \cite{Oberlack93} 
in agreement with the experimental data of the literature.

\bigskip

In general, Eq. (\ref{diffus}) represents models of diffusion approximation
based on the assumption that the turbulence can be represented by an opportune diffusivity which varies with $r$ \cite{Batchelor53}.
As a consequence of Eq. (\ref{diffus}),
$K$ contains  a term proportional to $\partial^2 f/ \partial r^2$ which can occur
only if the inertia forces include stochastic external terms, independent from the fluid state variables \cite{Skorokhod} and not present in the classical formulation \cite{Karman38}, \cite{Batchelor53}. 
For this reason the models based on Eq. (\ref{diffus}) are really phenomenological 
closure of Eq. (\ref{vk}).
\bigskip 

Although several other works on the von K\'arm\'an-Howarth equation were written \cite{Mellor84},
\cite{Onufriev94}, \cite{Grebenev05}, \cite{Grebenev09}, to the author's knowledge, a theoretical analysis based on basic principles which provides a physical-mathematical closure of the von K\'arm\'an-Howarth equation and the statistics of $\Delta u_r$ has not received due attention.
Therefore, the objective of the present work is to develop a theoretical analysis based on reasonable physical conjectures which allows the closure of the von K\'arm\'an-Howarth equation and the determination of the statistics of $\Delta u_r$.

Of course, besides the von K\'arm\'an-Howarth equation, there are
some other approaches, such as the
direct numerical simulation based on solving the Navier-Stokes
equations and the experiments, which are not considered in the present work.

\bigskip

The present work only considers the possibility to obtain the fully
developed homogeneous-isotropic turbulence in a given condition and
does not analyze the intermediate stages of the turbulence.
The study assumes that the fluctuations of the fluid state variables
are the result of the bifurcations of the Navier-Stokes equations. 
In section \ref{s2}, we present a qualitative scenario of these bifurcations which leads
to the onset of the turbulence. These bifurcations, defined by means of 
the fixed points of the velocity field and the Navier-Stokes equations, 
allows a rough estimation of the critical Reynolds number based on the Taylor scale.
After, in section \ref{s3},  the velocity fluctuation is studied through 
the kinematics of the local deformation and the momentum equations. 
These latter are expressed with respect to the referential coordinates which coincide with the material coordinates for a given fluid configuration \cite{Truesdell77}, whereas the kinematics of the local deformation is analyzed with the Lyapunov theory.
The choice of the referential coordinates allows the velocity fluctuations to be analytically expressed in terms of the Lyapunov exponent of the local fluid deformation.
The section \ref{s4} deals with the study of the velocity difference
between two fixed points of the space. This is analyzed with an opportune finite scale Lyapunov theory studying the motion of the particles crossing the two points, in the finite scale Lyapunov basis.
This basis is formally obtained through the orthonormalization method of Gram-Schmidt
of the finite scale Lyapunov vectors. 
These latter, also known as Bred vectors, were first introduced by Toth and Kalnay \cite{Toth93} to study the evolution in the time of a nonlinear perturbed model subjected to an initial finite perturbation. The choice of such vectors, whose properties are related to
the classical Lyapunov vectors \cite{Kalnay02}, is revealed to be an usefull tool for representing 
the relative motion of two particles crossing two given points of the space.

\no The present analysis postulates that the motion of such Lyapunov basis and that of the fluid with respect to the same basis, are completely statistically uncorrelated. 
This crucial assumption arises from the condition of fully developed turbulence.
The study leads to the closure of the von K\'arm\'an-Howarth equation \cite{Karman38} and gives an explanation of the mechanism of the kinetic energy transfer between length scales. 

The obtained expression of $K$ is the result of this assumption and does not correspond to
a diffusive approximation with model free parameters.
Its mathematical structure is in terms of $f$ and $\partial f / \partial r$ 
and satisfies the conservation law which states that the inertia forces only transfer the kinetic energy \cite{Karman38}, \cite{Batchelor53}.
This expression of $K$ corresponds to a first order term which makes the closed von K\'arm\'an-Howarth equation, a nonlinear partial differential equation of the first 
order in $r$ when $\nu =0$.
The main asset of the proposed closure on the other models is that it has been derived from a
specific Lyapunov theory, with reasonable basic assumptions about the statistics of the velocity
 difference. 

Furthermore, the statistics of the velocity difference is studied in section \ref{s7} with the Fourier analysis of the velocity fluctuations, and an analytical expression for the velocity difference and for its PDF is obtained in case of isotropic turbulence. This expression incorporates an unknown function, related to the skewness, which is identified through the obtained expression of $K$.
This velocity difference also requires the knowledge of the critical Reynolds number whose estimation is made in the section \ref{s2}. 

Finally, the several results obtained with this analysis are compared with the data
existing in the literature, indicating that the proposed analysis adequately describes
the various properties of the fully developed turbulence.


\section{\bf Bifurcations  \label{s2}}

This section qualitatively describes the route toward the
turbulence by means of the bifurcations of the Navier-Stokes equations
and provides an estimation of the critical Taylor scale Reynolds number,
assuming that the turbulence is fully developed, homogeneous and isotropic. 

The velocity field 
$
\ds {\bf u}  = {\bf u}({\bf x}, t)
\label{0_00}
\label{0_0}
$
of a viscous and incompressible fluid, measured in the reference frame $\Re$, 
satisfies the Navier-Stokes equations
\bea
\begin{array}{l@{\hspace{-1.cm}}l}
\nabla^* \cdot {\bf u}^* = 0 \\\\
\ds \frac{\partial {\bf u}^*}{\partial t^*}= - \left(  {\bf u}^* \nabla^* {\bf u}^* 
\ds + \nabla^* p^* - \frac{1}{Re}  {\nabla^*}^2 {\bf u}^* \right) 
\end{array}
\label{N-Sv1}
\eea
Into Eq. (\ref{N-Sv1}), ${\bf u}^* = {\bf u}/ U$,
$t^*= t U/L$, 
${\bf x}^* = {\bf x}/ L$, 
$p^* = p/ \rho U^2$, 
and $Re = U L/ \nu$, where $U$ and $L$ are assigned velocity and length,
respectively.
The pressure $p$ can be eliminated by taking the divergence of the momentum equation \cite{Batchelor53}.
The velocity field, starting
from the unique initial condition ${\bf u}({\bf x}, 0)$ which does not depend on the 
Reynolds number, will depend upon $Re$ by means of its time evolution
\bea
\ds \frac{\partial {\bf u}^*}{\partial t^*} = {\bf F}({\bf x}^*, t^*; Re)
\label{NS_res}
\eea
where ${\bf F}({\bf x}^*, t^*; Re)$ represents the right-hand-side of the 
momentum Navier-Stokes equations calculated for 
$\ds {\bf u}^*  = {\bf u}^*({\bf x}^*, t^*)$.

Consider now the velocity field at $t=0$,  
and the fixed points ${\bf X}$ of Eq. (\ref{NS_res}) which, by definition, satisfy 
$\ds {\partial {\bf u}^*}/ \partial t^*$ = 0 \cite{Guckenheimer90}.
Increasing the Reynolds number, ${\bf X}$ will vary according to Eq. (\ref{NS_res}),
which can be expressed through the implicit function theorem \cite{Guckenheimer90}
\bea
{\bf X}  = {\bf X}_0 - \int_{Re_0}^{Re} 
\nabla {\bf F}^{-1}
\frac{\partial{\bf F}  }{\partial Re} \ dRe
\label{bif_map}
\eea
where $Re$ plays the role of the control parameter and ${\bf X}_0$ is the fixed point calculated at $Re = Re_0$, for $t=0$.
The location of these points will depend on the momentum Navier-Stokes equations and
on the mathematical structure $\ds {\bf u}({\bf x}, 0)$ \cite{Guckenheimer90}.
Therefore, $Re$ influences the distribution of $\bf X$ in the space. 

According to the literature \cite{Feigenbaum78},  \cite{Ruelle71}, \cite{Pomeau80}, \cite{Eckmann81} and to the characteristics of the diverse kinds of bifurcations,
we assume the following qualitative scenario:

For small $Re$, the viscosity forces are stronger than the inertia ones 
and make $\bf F$ an almost smooth function of $\bf X$. 
When the Reynolds number increases, as long as the Jacobian 
$\nabla {\bf F}$ is nonsingular, $\bf X$ exhibits smooth
variations with $Re$, whereas at a certain $Re$, this Jacobian becomes singular
($\det \left( \nabla {\bf F}\right) $ = 0). 
This can correspond to the first bifurcation,
where at least one of the eigenvalues of 
$\nabla {\bf F}$ crosses the imaginary axis and $\bf X$ appears  to be discontinuous with
respect to $Re$ \cite{Guckenheimer90}.

\begin{figure}[b]
\centering
\vspace{-0.mm}
\hspace{-0.mm}
\includegraphics[width=0.450\textwidth]{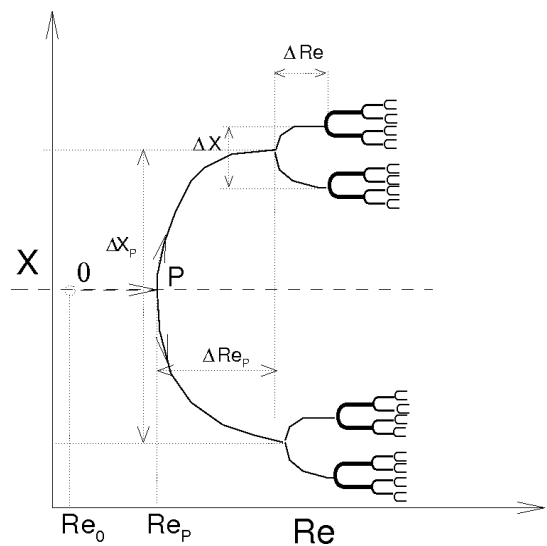}
\caption{Map of the bifurcations at an assigned time.}
\label{figura_1}
\end{figure}

Figure \ref{figura_1} shows a scheme of bifurcations at $t = 0$, where the component 
$X$ of $\bf X$ is reported in terms of $Re$. 
Starting from $Re_0$, the diagram is regular, until $Re_P$, where the first bifurcation
determines two branches whose maximum distance is $\Delta X_P$.
$\Delta X$ and $\Delta Re$ give, respectively, a length scale of the velocity field at the current value of $Re$, and the distance between two successive bifurcations.
After $P$, Eqs.  (\ref{NS_res}) and (\ref{bif_map}) do not indicate which of the two  branches the system will choose, thus a bifurcation causes a lost of informations with respect to the initial data \cite{Prigogine94}. 
That is, very small variations on the initial condition or very little perturbations, 
are of paramount importance for the choice of the branch that the system will follow \cite{Prigogine94}.
\
This is the situation of the bifurcations at $t = 0$. 

As the time increases, the position of the fixed points vary and so also the bifurcations, and far from the initial condition, one observes a developed motion, where the bifurcations can 
continuously vary with the time. 
Therefore, the bifurcations map changes with the time, and, if $Re$ is high enough, 
the length scales are continuously distributed. 
In this case the energy spectrum can be continuous and, according to the
theory \cite{Mathieu}, the velocity behaves as a chaotic function of $t$
and $\bf x$. There, the scales of Taylor and Kolmogorov are supposed to be assigned steady
quantities.

To describe the road to turbulence, observe that
the average distances $l_n \equiv \langle \Delta X_n \rangle$ depend
on $Re$ through Eqs. (\ref{NS_res}), and are here approximated by \cite{Guckenheimer90}
\bea
l_n = \frac{l_{1}} {\alpha^{n-1}}
\label{scales0}
\eea
where $\alpha$ $\approx 2$, \cite{Feigenbaum78} and the average is calculated on the time.
Equation (\ref{scales0}) is supposed to describe the route toward the chaos and is assumed to be valid until the onset of the turbulence. 
There, the minimum for $l_n$ can not be less than the Kolmogorov scale 
$\ds \ell = (\nu^3/ \varepsilon)^{1/4}$ \cite{Landau44}, \cite{Mathieu} where $l_1$ gives a good estimation of the correlation length of the phenomenon \cite{Guckenheimer90}, \cite{Prigogine94} which, in this case is the Taylor scale $\lambda_T$.
Thus, $\ell < l_n < \lambda_T$, and
\bea
\ds \ell = \frac{\lambda_T}  {\alpha^{N-1}}
\label{scales01}
\eea
where $N$ is the number of bifurcations at the beginning of the turbulence.
Equation (\ref{scales01}) gives the connection between the critical Reynolds number and $N$.
In fact, the characteristic Reynolds numbers associated to the scales 
$\ell$ and $\lambda_T$ are $R_K = \ell u_K/\nu \equiv$ 1 and  $R_{\lambda} = \lambda_T u/\nu$, respectively, where $\ds u_K = (\nu \varepsilon )^{1/4}$ is characteristic velocity at the
Kolmogorov scale,  and $u =\sqrt{\left\langle u_i u_i \right\rangle/3}$ \cite{Batchelor53}.
For isotropic turbulence, these scales are linked each other by \cite{Batchelor53}
\bea
\ds {\lambda_T}/{\ell} = 15^{1/4} \sqrt{R_\lambda}
\label{scales}
\eea
In view of Eq. (\ref{scales01}), this ratio can be also expressed through $N$, i.e.
\bea
\alpha^{N-1} = 15^{1/4} \sqrt{R_\lambda}
\label{scales1}
\eea
Assuming that $\alpha$ is equal to the Feigenbaum constant ($2.502...$),
the value $R_\lambda \simeq$ 1.6 obtained for $N=$ 2 is not compatible with $\lambda_T$
which is the correlation scale, while the result $R_\lambda \simeq$ 10.12, calculated for
$N=$ 3, is an acceptable minimum value for $R_\lambda$. The order of magnitude of these values 
can be considered in agreement with the various scenarios describing the roads to the turbulence \cite{Ruelle71}, \cite{Feigenbaum78}, \cite{Pomeau80}, \cite{Eckmann81}, and with the diverse 
experiments \cite{Gollub75}, \cite{Giglio81}, \cite{Maurer79} which state that the turbulence begins 
for $N \ge 3$.
Of course, this minimum value for $R_\lambda$ is the result of the assumptions 
$\alpha \simeq$ 2.502, $l_1 \simeq \lambda_T$, $l_N \simeq \ell$ and of 
approximation (\ref{scales0}).

\section{\bf Lyapunov analysis of the velocity fluctuations \label{s3}}

In this section, the velocity fluctuations caused by the bifurcations of 
Eqs. (\ref{N-Sv1}) \cite{Landau44},
are studied through the Lyapunov analysis of the kinematic of the fluid strain, using the Navier-Stokes equations.

Starting from the momentum Navier-Stokes equations written in a frame of reference $\Re$
\bea
\ds \frac{\partial  {u}_k}{\partial t}  = 
- \frac{\partial  {u}_k}{\partial x_h} u_h +
\frac{1}{\rho}  \frac{\partial T_{k h}}  {\partial x_{h}} 
\label{N-S}
\eea
consider the map $\bfchi$ : ${\bf x}_0 \rightarrow {\bf x}$, 
which is the function that determines the current position $\bf x$ of a fluid particle 
located at the referential position ${\bf x}_0$ \cite{Truesdell77} at $t = t_0$.
Equation (\ref{N-S}) can be written in terms of the referential position ${\bf x}_0$ \cite{Truesdell77} 
\bea
\ds \frac{\partial  {u}_k}{\partial t}  =  \left( -\frac{\partial  {u}_k}{\partial x_{0 p}} u_h +
\frac{1}{\rho}
 \frac{\partial T_{k h}}  {\partial x_{0 p}} \right)  \ \frac{\partial x_{0 p}}{\partial x_{h}} 
\label{N-Sr}
\eea
where $T_{k h}$ represents the stress tensor.
The Lyapunov analysis of the fluid strain provides the expression of this deformation in terms of the maximal Lyapunov exponent 
\bea
\frac{\partial {\bf x}}{\partial {\bf x}_0} \approx {\mbox e}^{\Lambda (t - t_0)} 
\label{stretch}
\eea
where $\Lambda = \max (\Lambda_1, \Lambda_2, \Lambda_3 )$ is the maximal Lyapunov exponent and $\Lambda_i$, $(i = 1, 2, 3)$ are the Lyapunov exponents.
Due to the incompressibility, $\Lambda_1 + \Lambda_2 + \Lambda_3 =$ 0,
thus,  $\Lambda >0$.

The momentum equations written using the referential coordinates allow
the factorization of the velocity fluctuation and to express it in Lyapunov exponential form of the local fluid deformation.
If we assume that this deformation is much more rapid than 
$ {\partial T_{k h}} / {\partial x_{0 p}}$ and ${\partial  {u}_k}/{\partial x_{0 p}} u_h$,
the velocity fluctuation can be obtained from 
Eq. (\ref{N-Sr}), where ${\partial T_{k h}} / {\partial x_{0 p}}$ and
${\partial  {u}_k}/{\partial x_{0 p}} u_h$ are supposed to be constant
with respect to the time 
\bea
\begin{array}{l@{\hspace{0cm}}l}
\ds u_k \approx 
\frac{1}{\Lambda}  \ 
\left( -\frac{\partial  {u}_k}{\partial x_{0 p}} u_h +
\frac{1}{\rho}
 \frac{\partial T_{k h}}  {\partial x_{0 p}} \right)_{t = t_0} 
 \ds \approx  \frac{1}{\Lambda} \left(  \frac{\partial u_k} {\partial t} \right)_{t = t_0} 
\end{array}
\label{fluc_v2_0}
\label{fluc_v2}
\eea
This assumption is justified by the fact that, according to Truesdell \cite{Truesdell77},  
${\partial T_{k h}} / {\partial x_{0 p}} -{\partial  {u}_k}/{\partial x_{0 p}} u_h$ is a smooth function of $t$ -at least during the period of a fluctuation-  whereas the fluid deformation varies very rapidly in the vicinity of a bifurcation according to Eq. (\ref{stretch}). This implies that, in proximity of  bifurcations, the Lyapunov basis of orthonormal vectors $E_\Lambda \equiv({\bf e}'_1, {\bf e}'_2, {\bf e}'_3)$ \cite{JiaLua05} associated to the strain (\ref{stretch}) rotates very quickly with respect to $\Re$ with an angular velocity $\bfomega_\Lambda$, whereas the modulus of the fluid velocity fluctuation, measured in $E_\Lambda$ increases with a rate $\approx {\mbox e}^{\Lambda (t - t_0)}$.
Since $\Lambda$ is related to the maximal eigenvalue of 
$(\nabla {\bf u} + \nabla {\bf u}^T )/2$,
according to the fluid kinematics  \cite{Ottino89}, \cite {Ottino90}, \cite{Lamb45}
 $\vert \bfomega_\Lambda \vert = O (\Lambda)$.
Therefore, the fluid velocity fluctuation measured in $\Re$, varies very quickly
because of the combined effect of the exponential growth rate and of the
rotations of $E_\Lambda$ with respect to $\Re$.

Note that, since $\partial {\bf x} / \partial {\bf x}_0$ is supposed to vary much more fastly
than the fluid state variables, as long as $t-t_0$ does not exceed very much the 
Lyapunov time $1/\Lambda$, the Lyapunov vectors almost coincide with the eigenvectors 
of $\nabla {\bf u}$ and $\Lambda$ is nearly the maximum eigenvalue of 
$(\nabla {\bf u} + \nabla{\bf u}^T )/2$ \cite{Guckenheimer90}.

\section{\bf Lyapunov analysis of the relative motion  \label{s4}}

\begin{figure}
\centering
\includegraphics[width=0.45\textwidth]{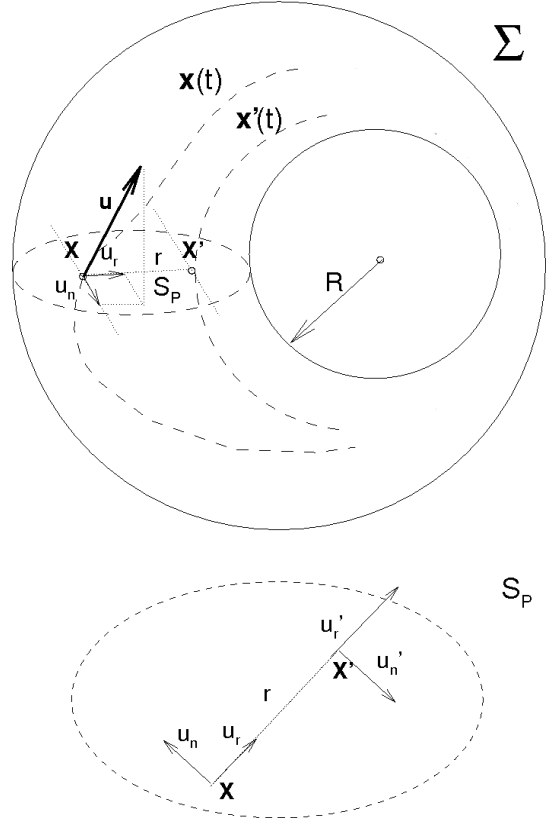}
\caption{Scheme of the relative motion of two fluid particles}
\label{figura_1a}
\end{figure}

In order to investigate the mechanism of the energy cascade, the relative motion between two fluid particles is here studied with the Lyapunov analysis.

Consider now two fixed points of the space, ${\bf X}$ and ${\bf X}'$ (see Fig. \ref{figura_1a}) with ${\bf r} = {\bf X}'-{\bf X}$, where $r = \vert {\bf r}\vert $
is the separation distance, and the motion of the two fluid particles which at the time $t_0$, cross through ${\bf X}$ and ${\bf X}'$. 
The equations of motion of these particles are
\bea
\begin{array}{l@{\hspace{+0.2cm}}l}
\ds \frac{d {\bf x}}{d t} = {\bf u} ({\bf x}, {t}),  \ \  \frac{d {\bf x}'}{d t}
 = {\bf u} ({\bf x}', {t})
\end{array}
\label{k_1}
\label{k_0}
\eea 
where ${\bf u} ({\bf x}, {t})$ and ${\bf u} ({\bf x}', {t})$ vary with the time according to
the Navier-Stokes equations.
$\hat{\bf r} = {\bf x}' -{\bf x}$ is the relative position between 
the two particles, therefore $\hat{\bf r} (t_0) = {\bf r}$.

The Lyapunov analysis of Eqs. (\ref{k_0}) leads to the determination of 
the maximal finite scale Lyapunov exponent $\lambda$ of
Eqs. (\ref{N-Sv1}) and (\ref{k_0})
\bea
\lambda(r) = \lim_{T \rightarrow \infty} \frac{1}{T} \int_0^T 
\frac{1} {\hat{\bf r}  \cdot \hat{\bf r}} \ 
\frac{d \hat{\bf r}}{dt}
 \cdot \hat{\bf r}
\ dt 
\label{Lyap_exp}
\eea
where $\lambda(0) = \Lambda$, and $r$ represents the scale associated to $\lambda$.
To define the finite scale Lyapunov basis of Eqs. (\ref{k_1}), 
consider the system 
\bea
\begin{array}{l@{\hspace{+0.2cm}}l}
\ds \frac{d {\bf x}}{d t} =  {\bf u} ({\bf x}, t)
\end{array}
\label{lyap_vector1}
\eea
\bea
\begin{array}{l@{\hspace{+0.2cm}}l}
\ds \frac{d \hat{\bf r}}{d t} =  {\bf u} ({\bf x} + \hat{\bf r}, t) - 
{\bf u} ({\bf x}, t) 
\end{array}
\label{lyap_vector2}
\eea
The finite scale Lyapunov vectors ${\bf r}_1$, ${\bf r}_2$ and ${\bf r}_3$ are defined as the solutions of Eq. (\ref{lyap_vector2}), starting from a given initial condition ${\bf r}_i(0)$, ($i$ =1, 2, 3) \cite{Toth93}, where ${\bf x}$ and $\bf u$ vary according to
Eqs. (\ref{lyap_vector1}) and  (\ref{N-Sv1}), respectively.
Such vectors, also known as Bred vectors, are, by definition, related to the classical Lyapunov vectors
in such a way that, ${\bf r}_1$, ${\bf r}_2$ and ${\bf r}_3$ tend to the classical Lyapunov vectors when   ${\bf r}_i(0) \rightarrow 0$, ($i$ =1, 2, 3) \cite{Kalnay02}, \cite{Annan04}.
The finite scale Lyapunov basis 
$E_\lambda \equiv({\bf e}_1, {\bf e}_2, {\bf e}_3)$ of Eqs. (\ref{k_1}) 
is then obtained by orthogonalizing ${\bf r}_1$, ${\bf r}_2$ and ${\bf r}_3$ 
with the Gram-Schmidt method at each time \cite{JiaLua05}, \cite{Annan04}. 

This basis rotates with respect to $\Re$ with an angular velocity $-\bfomega_\lambda$, 
where, because of isotropy $\vert \bfomega_\lambda \vert \approx \lambda(r)$.
The fluid velocity difference $\Delta {\bf v} \equiv (v_1' -v_1, \ v_2' -v_2, \ v_3' -v_3 )$,  measured in $E_\lambda$, is expressed by the Lyapunov theory
as
\bea 
\begin{array}{l@{\hspace{+0.0cm}}l}
\ds v_l' -v_l   =  {\lambda}_l \ \hat{r}_l, \ \ \ l = 1, 2, 3 
\end{array}
\label{Lyap}
\eea
Into Eq. (\ref{Lyap}), $\lambda_l$ is the Lyapunov exponent associated to the direction $\hat{r}_l$, where $\hat{r}_l$,  $v_l$ and  $v_l'$ are, respectively, the components of
${\bf x}'-{\bf x}$ and of ${\bf u} ({\bf X}, t) $ and 
${\bf u} ({\bf X}', t)$  measured in $E_\lambda$.
$\Delta {\bf v}$ can be also written in terms of the maximal exponent $\lambda$ 
\bea
\Delta {\bf v} (\hat{\bf r}) = \lambda(r) \hat{\bf r}  + {\bfzeta}
\eea 
where ${\bfzeta}$, due to the other two exponents, is negligible 
with respect to $\lambda {\bf r}$ and makes $\Delta {\bf v} (r)$ a solenoidal field, 
resulting  $\vert {\bfzeta} \vert << \vert \lambda {\bf r} \vert$ during the fluctuation.
When ${\bfzeta}/ \lambda r \rightarrow 0$, 
$\hat{\bf r}$ maintains unchanged orientation with respect to $E_\lambda$.
The velocity difference $\Delta {\bf u}$, measured in $\Re$ and expressed in 
$E_\lambda$, is calculated when $\vert \hat{\bf r} \vert = r$
\bea
\Delta {\bf u} (\hat{\bf r}) = \lambda(r) \hat{\bf r}  + {\bfzeta} + {\bfomega}_\lambda \times \hat{\bf r}
\label{LyapQ}
\eea 
This equation states that $\bfomega_\lambda$ determines the lateral components of the velocity difference in $E_\lambda$.

\bigskip

The longitudinal component of the velocity difference in $\Re$, $\Delta u_r$ is then
\bea
\Delta u_r (\hat{\bf r}) = {\bfxi} \cdot {\bf Q} \Delta {\bf u} (\hat{\bf r})
\label{v_re}
\eea
where ${\bf Q} \equiv ((e_{i j}))$ is the rotation matrix transformation from 
$E_\lambda$ to $\Re$,  $e_{i j}$ are the components of ${\bf e}_i$ in $\Re$,
and $\ds {\bfxi} = ({\bf X}'-{\bf X})/\vert {\bf X}'-{\bf X} \vert$ is the unit vector along the longitudinal direction. 
Observe that, $\lambda(r)$ and $\bfomega_\lambda$ arise both from the Navier-Stokes equations and that, in line with Eq. (\ref{Lyap_exp}), $\lambda(r)$ can change but its variations are much slower than $\bfomega_\lambda$ \cite{Guckenheimer90}.
Moreover, the Lyapunov vectors sweep a subdomain of their space of state which coincides with the physical space $R^3$. 
The simultaneous variations of these quantities produce the velocity difference fluctuation observed in $\Re$, according to Eq. (\ref{v_re}).

\bigskip

Although $\bfomega_\lambda$ and $\Delta {\bf v}$ arise from 
Eqs. (\ref{lyap_vector1}) and (\ref{lyap_vector2}), 
these are not directly related, therefore, we suppose here that $\bfomega_\lambda$ is statistically orthogonal to both ${\bf v}$ and 
${\bf v}'$, with $\langle \bfomega_\lambda \rangle =0$. 
This is the crucial assumption of the present work which is justified by the condition of fully developed turbulence. This provides that
\bea
\langle \bfomega_\lambda  v_p^m v_q'^n  \rangle = 0, \ m, n >0, \ p, q =1,2,3
\label{orth}
\eea
where the angular brackets denote the average calculated on the statistical ensemble of all the pairs of particles which cross through ${\bf X}$ and ${\bf X}'$.
Again thanks to the fully developed turbulence, we also suppose that 
$\bf Q$ and $\bfomega_\lambda$ are statistically orthogonal each other. This implies that
\bea
\langle \omega_{\lambda p}^m e_{i q}^n  \rangle = 
\langle \omega_{\lambda p}^m  \rangle \langle  e_{i q}^n  \rangle,  \ m, n > 0, \ i, p, q =1,2,3
\label{orth2}
\eea
whereas the isotropy gives
\bea
\langle \omega_{\lambda h} \omega_{\lambda k} \rangle = \frac{1}{3}
\langle \bfomega_\lambda   \cdot \bfomega_\lambda \rangle \delta_{h k}, \ \ 
\langle e_{h p} e_{k q} \rangle = \frac{1}{3}
 \delta_{h k}  \delta_{p q}
\label{isotropy} 
\eea
where $\bfomega_\lambda \equiv (\omega_{\lambda 1}, \omega_{\lambda 2}, \omega_{\lambda 3})$,
and ${\bf e}_i \equiv (e_{i 1}, e_{i 2}, e_{i 3})$, ($i = 1, 2, 3$), are expressed in $\Re$, and 
\bea
\langle \omega_{\lambda i}^2 \rangle = a_i \lambda^2(r), \ i = 1, 2, 3 
\label{isotropy_omega}
\eea
Because of homogeneity, $a_i = O (1)$ do not depend on $r$.

\bigskip

Under these hypotheses, we now show the following equations
\bea
\left\langle u_r  u_r' \omega_{\lambda k} \right\rangle
= 0  
\label{c1}
\eea
\bea
\left\langle u_n  u_n' \omega_{\lambda k} \right\rangle =
\left\langle u_b  u_b' \omega_{\lambda k} \right\rangle =
C_k u^2 \lambda (r) g(r)
\label{c2}
\eea 
where $u_r$, $u_n$ and $u_b$ are the components of ${\bf u}$ along $\hat{\bf r}$ and along the two  orthogonal directions $n$ and $b$, and $C_k$ is a proper constant of the order of unity.
The function $g(r) = \langle u_n u_n' \rangle/u^2$ is the lateral velocity correlation function (see Appendix) measured in $E_\lambda$, that due to the homogeneity and isotropy, coincides with
the lateral correlation function measured in $\Re$.

Equation (\ref{c1}) is the direct consequence of Eq. (\ref{orth}).
In fact $\langle u_r u_r' \bfomega_\lambda \rangle $ 
$\equiv$ $\langle v_r v_r' \bfomega_\lambda \rangle = 0$.

To demonstrate Eq. (\ref{c2}), observe that $u_n$ and $\bfomega_\lambda$ can be decomposed into
$n$ independent,  identically distributed, stochastic variables $\xi_k$, 
which satisfy \cite{Lehmann99}
\bea
\langle \xi_k \rangle = 0, \ \ \ 
\langle \xi_h \xi_k \rangle = \delta_{h k} \ \ \
\langle \xi_h \xi_k \xi_l \rangle =  \varpi_{h k j} q
\label{deltaK}
\eea
where $\varpi_{h k j} = 1$ if $h = k = j$, else  $\varpi_{h k j} = 0$ and $q \ne 0$.
$\omega_{\lambda j}$ is expressed as the sum of $\xi_k$, where, without loss of generality, 
the coefficients of the combination are assumed constant with respect to $\bf X$ and equal each other \cite{Lehmann99}.
\bea
\omega_{\lambda j} = A_j \lambda(r) \frac{1}{n} \sum_{k= 1}^n \xi_k 
\label{om1}
\eea 
with $A_1 = A_2 = A_3 = O(1)$ constant parameters.
The component $u_n$ (and $u_b$)  
is also expressed as the linear combination of $\xi_k$, whose coefficients are now functions of $\bf X$ as the consequence of the previous assumption, i.e.
\bea
u_n ({\bf X}) = u \sum_{k= 1}^n F_k({\bf X}) \xi_k
\label{u_n}
\eea
Hence, the correlation function $g(r)$ is in terms of $F_k$ and is determined through 
Eq. (\ref{u_n}) and (\ref{deltaK}) \cite{Lehmann99}, putting ${\bf X}=0$.
\bea
g(r) = \sum_{k = 1}^n F_k(0) F_k({\bf r})
\label{gr} 
\eea
Therefore, $\left\langle u_n  u_n' \omega_{\lambda k} \right\rangle$ is calculated taking 
into account Eq. (\ref{om1}), (\ref{u_n}), (\ref{gr}) and (\ref{deltaK}), and, as a result,
 Eq. (\ref{c2}) is achieved.

\section{\bf Closure of the von K\'arm\'an-Howarth equation  \label{s5}}

Now, we present the closure of the von K\'arm\'an-Howarth equation based on the analysis
seen at the previous section.

The term representing the inertia forces in the von K\'arm\'an-Howarth equation
satisfies the identity (\ref{vk1}) (see Appendix), which is here written as 
\cite{Karman38}, \cite{Batchelor53}
\bea
\frac{\partial}{\partial {r}_k}  \left( {r}_k K \right) =
\ds  \frac{\partial }{\partial {r}_k} 
\ds  \left\langle u_i  u_i' (u_k - u_k')  \right\rangle 
\label{Tr}  
\eea
where, due to the isotropy $K$ is a function of $r$ alone \cite{Karman38}. 
The divergence of ${\bf r} K$ gives the mechanism of the energy cascade 
 which does not depend upon the frame of reference
\cite{Karman38} \cite{Batchelor53}.
In order to determine the expression of  $K$, Eq. (\ref{Tr}) is here written in $E_\lambda$. In view of Eq. (\ref{LyapQ}) and taking into account that $u_i u_i'$ is also frame independent, one obtains the following equation
\bea
\frac{\partial}{\partial \hat{r}_k}  \left(  \hat{r}_k K \right) =
\ds -  \frac{\partial }{\partial \hat{r}_k} \left( 
\ds  \left\langle u_i  u_i' \lambda \right\rangle \hat{r}_k  
+  (\left\langle  u_i  u_i'  {\bfomega}_\lambda  \right\rangle \times \hat{\bf r})_k
+ \left\langle u_i  u_i' \zeta_k \right\rangle \right) 
\label{Tr_L}  
\eea
Since $\lambda$ is calculated with Eq. (\ref{Lyap_exp}), this is constant with respect to the statistics of  $u_i$ and  $u_i'$, thus $\langle {\lambda} u_i u_i'  \rangle = {\lambda} \langle u_i u_i' \rangle$, and $K$ is expressed as the general integral of Eq. (\ref{Tr_L})
\bea
K \hat{\bf r} = - \lambda \left\langle u_i  u_i'  \right\rangle \hat{\bf r}  
- \left\langle u_i  u_i' {\bfomega}_\lambda \right\rangle \times  \hat{\bf r} 
+{\bf s}
\label{Kr}
\eea
Into Eq. (\ref{Kr}), ${\bf s}$ is the sum of a term due to $\bfzeta$ plus an arbitrary
solenoidal field arising from the integration of Eq. (\ref{Tr_L}) \cite{Vector}.
According to this analysis of Lyapunov, $\bf s$ is proportional to $u^2 \lambda r$ and
can be written in the form 
\bea
{\bf s} = u^2 \lambda(r) r \ {\bf s}_0
\label{ss}
\eea
Substituting Eq. (\ref{R_corr}) (see Appendix) and Eq. (\ref{ss})  into Eq. (\ref{Kr}), 
$K \hat{\bf r}$ is 
\bea
K \hat{\bf r} =  \left( \lambda u^2 (g -f) - 3 \lambda u^2 g \right)  \hat{\bf r}  
- \left\langle u_i  u_i' {\bfomega}_\lambda \right\rangle \times \hat{\bf r} 
+ u^2 \lambda(r) r \ {\bf s}_0
\label{Kr I}
\eea
where $f(r)$ is the longitudinal velocity correlation function.

Equation (\ref{Kr I}) is made by three addends. 
In the first one of these, the part into the brackets is an even function of $r$ which goes to zero as
$r \rightarrow \infty$ and assumes the value $-3 u^2 \lambda(0)$ for $r=0$.
The second term,  orthogonal to the first one, vanishes at $r=0$ and tends to zero
when $r \rightarrow \infty$.
In this latter $u_i u_i'$ is expressed in $E_\lambda$ for sake of convenience 
\bea
\left\langle u_i  u_i' {\bfomega}_\lambda \right\rangle =
\left\langle u_r  u_r' {\bfomega}_\lambda \right\rangle +
\left\langle u_n  u_n' {\bfomega}_\lambda \right\rangle+
\left\langle u_b  u_b' {\bfomega}_\lambda \right\rangle
\label{eq c}
\eea
The first term at the RHS of Eq. (\ref{eq c}) vanishes because of Eq. (\ref{c1}),
whereas the other ones are expressed by means of Eq. (\ref{c2}).
Therefore
\bea
\left\langle u_i  u_i' {\bfomega}_\lambda \right\rangle \times \hat{\bf r} =
2 u^2 \lambda(r) g(r) {\bf c} \times \hat{\bf r}
\eea
where ${\bf c} \equiv ( C_1, C_2, C_3)$ and $C_k =O(1)$ are from Eq. (\ref{c2}).

In order to satisfy Eq. (\ref{Kr I}), the expression of ${\bf s}_0$ must be of the kind 
$
{\bf s}_0 = h(r) \hat{\bf t} + p(r)  {\bf n}
$,
where, because of homogeneity and according to Refs. \cite{Batchelor53} and \cite{Robertson40}, $h(r)$ and $p(r)$ are even functions of $r$. 
Moreover, due to the isotropy and without lack of generality, $h(r) = p(r)$ 
\cite{Batchelor53},  \cite{Robertson40}, so  ${\bf s}_0$ is 
\bea
{\bf s}_0 =  h(r) ( {\bf t} + {\bf n} ) 
\eea
where ${\bf t}=\hat{\bf r}/r $, 
${\bf n} =({\bf c} \times \hat{\bf r})/\vert {\bf c} \times \hat{\bf r} \vert$.

To determine $K$ and $h(r)$, Eq. (\ref{Kr}) is projected along the directions 
$\hat{\bf r}$ and ${\bf n}$ 
\bea
\begin{array}{l@{\hspace{+0.0cm}}l}
\ds K =  \lambda u^2 (g-f)  -3 u^2 \lambda g + \frac{{\bf s} \cdot \hat{\bf r}}{r^2} 
\end{array}
\label{Kr1}
\eea
\bea
\begin{array}{l@{\hspace{+0.0cm}}l}
\ds  {\bf s} \cdot {\bf n} = 
\left\langle u_i  u_i' {\bfomega}_\lambda \times \hat{\bf r} \right\rangle \cdot {\bf n}
\end{array}
\label{Kr2}
\eea
From Eq. (\ref{Kr2}) 
\bea
\ds h(r) = H g(r)
\label{h-g}  
\eea
where
$
H = 2  ({\bf c} \times \hat{\bf r} )\cdot {\bf n}/{r} = O(1)
$.
As $\bfzeta/r \rightarrow 0$, $\hat{\bf r}$ 
maintains unchanged orientation in $E_\lambda$, thus
$H$ is an invariant which has to be identified.
The function $K$ is determined with Eq. (\ref{Kr1})
\bea
\begin{array}{l@{\hspace{+0.0cm}}l}
K =  \lambda u^2 (g-f)  + u^2 \lambda(r) g(r) (H-3)  
\end{array}
\label{Kr3}
\eea
$K(r)$  satisfies the conditions 
$\partial K(0) /\partial r = 0$ and $K(0) = 0$ \cite{Batchelor53},
which represent, respectively, the homogeneity of the flow and the condition that the inertia forces do not modify the fluid kinetic energy.
Since $g(0) = f(0) = 1$, this immediately identifies $H=3$  and
\bea
\begin{array}{l@{\hspace{+0.0cm}}l}
K =
\ds \lambda \ u^2 \ ( g - f ) 
\end{array}
\label{vv'}
\eea
Due to the fluid incompressibility, $f$ and $g$ are related each other through 
$g = f +1/2 \partial f /\partial r \ r$ (see Eq. (\ref{g}), Appendix), leading
to the expression
\bea
K =
\frac{1}{2} u^2  \frac{\partial f}{\partial r} \ \lambda(r) r
\label{vv'1}
\eea
This expression of $K$ has been obtained studying the properties of the velocity difference in $E_\lambda$.

Equation (\ref{vv'1}) states that, the fluid incompressibility, expressed by 
$g - f \ne 0$, 
represents a sufficient condition to state that $K \ne$ 0. 
This latter is determined as soon as $\lambda$ is known. 
To calculate $\lambda$, it is convenient to express 
$\Delta {\bf u} = {\bf u}({\bf x}', t) - {\bf u}({\bf x}, t)$ in $\Re$, with 
$\vert \hat{\bf r} \vert = r$.
Thus, $\Delta u_r$ is first expressed in terms of $\hat{\bf r}$ and $\Delta {\bf v}$ 
through Eq. (\ref{v_re}) ($\Delta u_r = {\bfxi} \cdot   {\bf Q} \Delta {\bf u}$),
then its standard deviation is calculated assuming that 
$\bfzeta =0$, 
(i.e. $\Delta {\bf u} = \lambda \hat{\bf r} + {\bfomega}_\lambda \times \hat{\bf r}$) 
and taking into account that $\bf Q$ and $\bfomega_\lambda$ satisfy
Eq. (\ref{isotropy}), with $\langle \bfomega_\lambda \rangle$ = 0:
\bea
\begin{array}{l@{\hspace{+0.0cm}}l}
\left\langle (\Delta u_r)^2 \right\rangle = 
\sum_{i, j, k, l} \sum_{p, q, r, s}  ( \xi_i \xi_p
\langle {\lambda}^2  e_{i j} e_{p q} \rangle \hat{r}_j \hat{r}_q +
\xi_i \xi_p \langle \lambda e_{i j} e_{p q} \omega_{\lambda r} \rangle \hat{r}_j \hat{r}_s \varepsilon_{q r s}+
\xi_i \xi_p \langle \lambda e_{i j} e_{p q} \omega_{\lambda k} \rangle \hat{r}_q \hat{r}_l \varepsilon_{j k l}+ \\\\
\xi_i \xi_p \langle e_{i j} e_{p q} \omega_{\lambda k} \omega_{\lambda r} \rangle \hat{r}_l \hat{r}_s ) \varepsilon_{j k l} \varepsilon_{q r s}
\end{array}
\label{1A_1}
\eea 
where 
$\bfxi \equiv  (\xi_1, \xi_2, \xi_3)$ and 
$\varepsilon_{i j k} = (j-i)(k-i)(k-j)/2$ represents the Levi-Civita tensor arising from 
the cross product.
\\
Since $\lambda$ is the average of the velocity increment per unit distance, it is constant with respect the statistics of $\bf Q$ and $\bfomega_\lambda$, thus
$\langle {\lambda}^2  ... \rangle = {\lambda}^2 \langle  ... \rangle$.
Because of Eq. (\ref{orth2}), $\bf Q$ and $\bfomega_\lambda$ are 
statistically independent, so that 
$\langle e_{i j} e_{p q} \omega_{\lambda k} \rangle =$ 
$\langle e_{i j} e_{p q} \rangle \langle \omega_{\lambda k} \rangle =$0 and
$ \langle e_{i j} e_{p q} \omega_{\lambda k} \omega_{\lambda r} \rangle =$
$ \langle e_{i j} e_{p q} \rangle \langle \omega_{\lambda k} \omega_{\lambda r} \rangle$, 
therefore second and third terms vanish into Eq. (\ref{1A_1}).
\\
Due to the isotropy, the Lyapunov basis satisfies Eq. (\ref{isotropy})  
($\langle e_{i j} e_{p q} \rangle = \delta_{i p} \delta_{j q}/3$). This  
is introduced in the first term of Eq. (\ref{1A_1}), which thus depends on $r^2$ alone. 
This term, which equals $\lambda^2 r^2/3$, is associated to the direction $\hat{\bf r}$ 
(maximal Lyapunov exponent direction) and represents one degree of freedom in the space. 
Again thanks to the isotropy, the last term of Eq. (\ref{1A_1}) is two times the first one because it is caused by $\bfomega_\lambda$ which corresponds to the two directions
orthogonal to $\hat{\bf r}$ and thus to two degrees of freedom.
As a result, taking into account that 
$\varepsilon_{i j k} \varepsilon_{i j h} = 2 \delta_{h k}$,
the standard deviation of the longitudinal velocity difference is
\bea
\left\langle (\Delta u_r)^2 \right\rangle = {\lambda}^2 r^2
\label{1A}
\eea 
with 
\bea
\ds \lambda(r) = 
\sqrt{\frac{\langle \bfomega_\lambda \cdot \bfomega_\lambda \rangle}{3}}
\eea
This standard deviation can be expressed through the longitudinal correlation function $f$
\bea
\langle (\Delta u_r)^2 \rangle = 2 u^2 (1-f(r))
\label{1B}
\eea
being $u$ the standard deviation of the longitudinal velocity.
The maximal Lyapunov exponent is calculated in function of $f$,
from Eqs. (\ref{1A}) and (\ref{1B})
\bea
\ds {\lambda} (r) = \frac{u}{r} \sqrt{2 \left( 1-f(r) \right) }
\label{lC}
\eea
Hence, substituting Eq. (\ref{lC}) into Eq. (\ref{vv'1}), one obtains
the expression of $K$ in terms of $f$ and its gradient
\bea
K =
u^3 \sqrt{\frac{1-f}{2}} \ \frac{\partial f}{\partial r} 
\label{vv'2}
\label{vk6}
\eea
This is the proposed closure of the von K\'arm\'an-Howarth equation, 
the main asset of which on the different diffusion models is that it has been
derived from a specific Lyapunov analysis, under the assumption of reasonable statistical hypotheses about the longitudinal velocity difference.
According to Eq. (\ref{vk6}), $K$ is a nonlinear term of the first order, 
thus it does not represent a diffusion approximation, but rather a nonlinear 
advection term which makes Eq. (\ref{vk}) a nonlinear partial differential equation
of the first order when $\nu =0$.

Equation (\ref{vk6}) corresponds to a mechanism of the kinetic energy transfer which preserves the average values of the momentum and of the  kinetic energy. Specifically, the analytical structure of Eq.(\ref{vk6}) states that this mechanism consists of a flow of the kinetic energy from large to small scales which only redistributes the kinetic energy between wavelengths.

This mechanism can be interpreted as follows.
If, at $t_0$, a toroidal volume $\Sigma (t_0)$ is taken which contains 
${\bf X}$ and ${\bf X}'$ (see Fig. \ref{figura_1a}), its geometry and position change according to the fluid motion, and its dimensions, $\sqrt{S_p}$ and $R$, vary with the time to preserve the volume.
Choosing $\Sigma$ in such a way that $R$ increases with the time, 
the Lyapunov analysis of Eqs. (\ref{k_0}) leads to  
$
\ds R \approx  R(t_0) \ \mbox{e}^{\lambda (t-t_0)}
\label{Lyapunov}
$.
According to the theory \cite{Guckenheimer90}, for $t > t_0$, the trajectories of the two particles are enclosed into $\Sigma (t)$.
Hence, the kinetic energy, initially enclosed into $\Sigma(t_0)$, at the end of the fluctuation is contained into $\Sigma(t)$ whose dimensions are changed with respect to $\Sigma(t_0)$. The kinetic energy is then transferred far from $\bf X$
and $\bf X'$, resulting enclosed in a more thin toroid.

\section{\bf Skewness of the velocity difference PDF \label{s6} }

The obtained expression of $K(r)$ allows to determine the skewness of $\Delta u_r$ \cite{Batchelor53}
\bea
\ds H_3(r) = \frac{\left\langle (\Delta u_r)^3 \right\rangle} 
{\left\langle (\Delta u_r)^2\right\rangle^{3/2}} =
  \frac{6 k(r)}{\left( 2 (1 -f(r)  )   \right)^{3/2} }
\label{H_3_01}
\eea
which is expressed in terms of the longitudinal triple correlation $k(r)$, linked to $K(r)$
by  $K(r)= u^3 \left( {\partial}/{\partial r}  + 4/r  \right) k(r)$ (also see Appendix, 
Eq. (\ref{kk})).
Since $f$ and $k$ are, respectively, even and odd functions of $r$ with
 $f(0)$ = 1, $k(0) = k'(0)=k''(0)$ =0,   $ H_3(0)$ is given by 
 \bea
\ds H_3(0) = \lim_{r\rightarrow0} H_3(r) = \frac{k'''(0)}{(-f''(0))^{3/2}} 
\label{H_3_0}
\eea
where the apex denote the derivative with respect to $r$.
To obtain $H_3(0)$, observe that, near the origin, $K$ behaves as
\bea
\begin{array}{l@{\hspace{+0.cm}}l}
 \ds K = u^3 \sqrt{-f''(0)} f''(0) \frac{r^2}{2} + O(r^4)
\end{array}
\label{K0}
\eea
then, substituting Eq. (\ref{K0}) into $K(r)= u^3 \left( {\partial}/{\partial r}  + 4/r  \right) k(r)$ and accounting for Eq. (\ref{H_3_0}), one obtains
\bea
\ds H_3(0)  = -\frac{3}{7} = -0.42857...
\label{sk0}
\eea
$H_3(0)$ is a constant of the present analysis, 
which does not depend on the Reynolds number. 
This is in agreement with the several sources of data existing in the literature
such as \cite{Batchelor53}, \cite{Tabeling96}, \cite{Tabeling97}, \cite{Antonia97} (and Refs. therein) and its  value gives the entity of the mechanism of the energy cascade.

This skewness causes variations in the time of the Taylor scale in accordance to Eq. (\ref{vk}).
The time derivative of $\lambda_T$ is calculated considering the coefficients 
of the order  $O (r^2)$ in the von K\'arm\'an-Howarth equation 
\cite{Karman38}, \cite{Batchelor53} which are determined substituting Eqs. (\ref{K0}) and (\ref{sk0}) into Eq. (\ref{vk})
\bea
\frac{d \lambda_T}{dt} = 
 -\frac{u}{2} + \nu \left( \frac{7}{3} f^{IV}(0) \lambda_T^3 + \frac{5}{\lambda_T} 
\right) 
\label{dl/dt}
\eea
where $f^{IV} \equiv \partial^4 f /\partial r^4$.
The first term tends to decrease $\lambda_T$ and expresses the mechanism of energy cascade, whereas the term in the brackets gives the viscosity effect which, in general, tends to increase $\lambda_T$ depending on the current values of $f^{IV}(0)$ and $\lambda_T$ \cite{Karman38}, \cite{Batchelor53}.

\section{\bf Statistical analysis of the velocity difference \label{s7}}

As explained in this section, the Lyapunov analysis of the local deformation
and some plausible assumptions about the statistics of the velocity difference
$\Delta {\bf u} ({\bf r}) \equiv {\bf u} ({\bf X}+{\bf r})-{\bf u} ({\bf X})$
lead to determine all the statistical moments of $\Delta {\bf u} ({\bf r})$
with only the knowledge of the function $K(r)$ and of the value of the critical
Reynolds number.

The statistical properties of $\Delta {\bf u} ({\bf r})$, are  investigated expressing the velocity fluctuation, given by Eq. (\ref{fluc_v2}), as the Fourier series
\bea
\ds {\bf u} 
\approx
 \frac{1}{\Lambda} \sum_{\bfkappa} \frac{{\bf \partial U}}{\partial t} ({\bfkappa}) 
 {\mbox e}^{i {\bfkappa}\cdot {\bf x}}  
\label{f1}
\eea
where ${\bf U}({\bfkappa})$ $\equiv$ $(U_1({\bfkappa}), U_2({\bfkappa}), U_3({\bfkappa}))$ are the components of the velocity spectrum, which satisfy the Fourier transformed Navier-Stokes equations \cite{Batchelor53}
\bea
\begin{array}{l@{\hspace{0.cm}}l}
\ds  \frac{\partial  U_p ({\bfkappa})}{\partial t}  = - \nu k^2 U_p ({\bfkappa}) + 
\ds i \sum_{\bf j} ( \frac{\kappa_p \kappa_q \kappa_r}{\kappa^2}   U_q({\bf j}) 
U_r({\bfkappa} -{\bf j}) 
\ds - \kappa_q  U_q({\bf j}) U_p({\bfkappa} -{\bf j}) ) 
\end{array}
\label{NS_fourier}
\eea
All the components ${\bf U}({\bfkappa}) \approx  \partial {\bf U}({\bfkappa})/ {\partial t} /\Lambda$ are random variables distributed according to certain distribution functions, which are statistically orthogonal each other \cite{Batchelor53}.

Thanks to the local isotropy, $\bf u$ is sum of several dependent random variables which are identically distributed \cite{Batchelor53}, therefore $\bf u$ tends to a gaussian variable \cite{Lehmann99}, and ${\bf U}({\bfkappa})$ satisfies the Lindeberg condition, a very general necessary and sufficient condition for satisfying the central limit theorem \cite{Lehmann99}. 
This condition does not apply to the Fourier coefficients of 
$\Delta {\bf u}$. In fact, since $\Delta {\bf u}$ is the difference between two dependent gaussian variables, its PDF could be a non gaussian
distribution function.
In ${\bf x}=0$, the velocity difference $\Delta {\bf u} ({\bf r}) \equiv
(\Delta u_1, \Delta u_2, \Delta u_3)$ is given by
\bea
\Delta u_p \hspace{-1.mm} \approx \hspace{-1.mm}   \frac{1}{\Lambda} 
\sum_{\bfkappa} \frac{\partial  U_p ({\bfkappa})} {\partial t} 
({\mbox e}^{i {\bfkappa}\cdot {\bf r}} - 1)    \equiv L + B + P + N
\eea
This fluctuation consists of the contributions appearing into Eq. (\ref{NS_fourier}):
in particular, $L$ represents the sum of all linear terms due to the viscosity
and $B$ is the sum of all bilinear terms arising from inertia and pressure
 forces. $P$ and $N$  are, respectively, the sums of definite positive 
and negative square terms, which derive from inertia and pressure forces.
The quantity $L+B$ tends to a gaussian random variable being 
the sum of statistically orthogonal terms \cite{Madow40}, \cite{Lehmann99}, while $P$ and $N$ do
 not, as they are linear combinations of squares \cite{Madow40}.
 Their general  expressions are  \cite{Madow40}
\bea
\begin{array}{l@{\hspace{+0.2cm}}l}
 P = P_0 + \eta_1  +  \eta_2^2   \\\\
 N = N_0 + \zeta_1 -  \zeta_2^2  
\end{array} 
\label{nn}
\eea
where $P_0$ and $N_0$ are constants, and $\eta_1$, $\eta_2$, $\zeta_1$ and  $\zeta_2$ are four different centered random gaussian variables. 
Therefore, the fluctuation $\Delta u_r$ of the longitudinal velocity difference
can be written as
\bea
\begin{array}{l@{\hspace{+0.2cm}}l}
\ds \Delta {u}_p  = 
\psi_1({\bf r}) {\xi} + \psi_2({\bf r})  
\ds \left( \chi  ( {\eta}^2-1 )  -  ( {\zeta}^2-1 )  \right) 
\end{array}
\label{fluc3}
\eea
where $\xi$, ${\eta}$ and $\zeta$ are independent centered random variables
which have gaussian distribution functions with standard deviation equal to the unity.
The parameter $\chi$ is a positive definite function of the Reynolds number, 
whereas $\psi_1$ and $\psi_2$ are functions of the space
coordinates and the Reynolds number. 

At the Kolmogorov scale $\ell$, the order of magnitude of the velocity fluctuations is ${u_K}^2 \tau/\ell$, with $\tau = 1/\Lambda$ and $u_K = \nu / \ell$, whereas $\psi_2$ is negligible because is due to the inertia forces: this immediately identifies $\psi_1 \approx {u_K}^2 \tau/\ell$.
\\
On the contrary, at the Taylor scale $\lambda_T$, $\psi_1$ is negligible and the order of magnitude of the velocity fluctuations is $u^2 \tau/\lambda_T$, therefore $\psi_2 \approx u^2 \tau/\lambda_T$.

The ratio $\psi_2 / \psi_1$ is a function of $R_\lambda$
\bea
\psi({\bf r}, R_{\lambda}) = \frac{\psi_2 ({\bf r})}{\psi_1({\bf r})} 
\approx \frac{u^2 \ell}{{u_K}^2 \lambda_T} =  
\sqrt{\frac{R_{\lambda}}{15 \sqrt{15}}} \
\hat{\psi}({\bf r})
\label{Rl}
\eea
where 
$
\ds \hat{\psi}({\bf r}) = O(1)
\label{R2}
$, 
is a function which has to be determined.

Hence, the dimensionless longitudinal velocity difference $\Delta {u}_r$, is written as
\bea
\begin{array}{l@{\hspace{+0.2cm}}l}
\ds \frac {\Delta {u}_r}{\sqrt{\langle (\Delta {u}_r)^2} \rangle} =
\ds \frac{   {\xi} + \psi \left( \chi ( {\eta}^2-1 )  -  
\ds  ( {\zeta}^2-1 )  \right) }
{\sqrt{1+2  \psi^2 \left( 1+ \chi^2 \right)} } 
\end{array}
\label{fluc4}
\eea 
The dimensionless statistical moments of $\Delta {u}_r$
are easily calculated considering that $\xi$, $\eta$ and 
$\zeta$ are independent gaussian variables
\bea
\begin{array}{l@{\hspace{+0.2cm}}l}
\ds H_n \equiv \frac{\left\langle (\Delta u_r)^n \right\rangle}
{\left\langle (\Delta  u_r)^2\right\rangle^{n/2} }
= 
\ds \frac{1} {(1+2  \psi^2 \left( 1+ \chi^2 \right))^{n/2}} 
\ds \sum_{k=0}^n 
\left(\begin{array}{c}
n  \\
k
\end{array}\right)  \psi^k
 \langle \xi^{n-k} \rangle 
  \langle (\chi(\eta^2 -1) - (\zeta^2 -1 ) )^k \rangle 
\end{array}
\label{m1}
\eea
where
\bea
\begin{array}{l@{\hspace{+0.2cm}}l}
\ds   \langle (\chi(\eta^2 -1) - (\zeta^2 -1 ) )^k \rangle = 
\ds \sum_{i=0}^k 
\left(\begin{array}{c}
k  \\
i
\end{array}\right)  
(-\chi)^i 
 \langle (\zeta^2 -1 )^i \rangle 
 \langle (\eta^2 -1 )^{k-i} \rangle \\\\
\ds  \langle (\eta^2 -1 )^{i} \rangle = 
\sum_{l=0}^i 
\left(\begin{array}{c}
i  \\
l
\end{array}\right)  
(-1)^{l}
\langle \eta^{2(i-l)} \rangle 
 \end{array}
\label{m2}
\eea
In particular, the third moment or skewness, $H_3$,
which is responsible for the energy cascade, is
\bea
\ds H_3= \frac{  8  \psi^3 \left( \chi^3 - 1 \right) }
 {\left( 1+2  \psi^2 \left( 1+ \chi^2 \right) \right)^{3/2}  }
\label{H_3}
\eea 
For $\chi \ne$ 1, the skewness and all the odd order moments are different from zero,
and for $n>3$, all the absolute moments are rising functions of $R_{\lambda}$, 
thus $\Delta u_r$ exhibits an intermittency whose entity increases with the Reynolds number.

\begin{figure}
\vspace{-0.mm}
\centering
\includegraphics[width=0.45\textwidth]{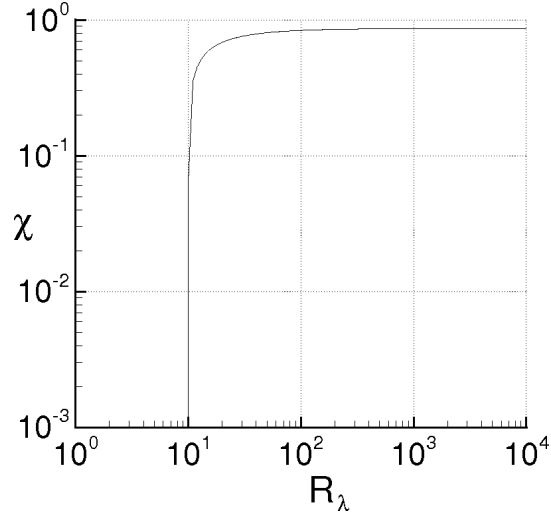}
\vspace{-2.mm}
\caption{Parameter $\chi$ plotted as the function of $R_{\lambda}$.}
\vspace{-0.mm}
\label{figura_2}
\end{figure}

All the statistical moments can be calculated once the function $\chi(R_\lambda)$
and the value of $\hat{\psi}_0$ are known.
The expression of $K(r)$ obtained in the first part of the work allows to
identify $H_3(0)$ and then fixes the relationship between $\psi_0$ and $\chi(R_\lambda)$
\bea
-H_3(0) = \frac{  8  {\psi_0}^3 \left(  1-\chi^3 \right) }
 {\left( 1+2  {\psi_0}^2 \left( 1+ \chi^2 \right) \right)^{3/2}  }
= \frac{3}{7}
\label{sk1}
\eea
where ${\psi}_0 = \psi(0, R_{\lambda})= O (\sqrt{R_{\lambda}})$ and 
$\chi = \chi(R_\lambda) > 0$.
This relationship does not admit solutions with $\chi > 0$ below a minimum value
of $(R_\lambda)_{min}$ dependent on $\hat{\psi}_0$.
According to the analysis of section \ref{s2}, $(R_\lambda)_{min}$
is chosen to 10.12, which corresponds to $\hat{\psi}_0 \simeq 1.075$.
(setting $\chi=0$, $R_\lambda$ = 10.12 in $H_3(0)$).
Varying the value of $(R_\lambda)_{min}$ from 8.5 to 15 would bring values of
$\hat{\psi}_0$ between 1.2 and 0.9, respectively.
In figure \ref{figura_2}, the function $\chi(R_\lambda)$ is shown for
 $\hat{\psi}_0 = 1.075$.
The limit  $\chi \simeq$ 0.86592 for $R_{\lambda} \rightarrow \infty$ is
reached independently of the value of $\hat{\psi}_0$.

The PDF of $\Delta u_r$ is expressed through the Frobenius-Perron equation
\bea
\begin{array}{l@{\hspace{+0.0cm}}l}
F(\Delta {u'}_r) = \hspace{-0.mm}
\ds \int_\xi \hspace{-0.mm}
\int_\eta  \hspace{-0.mm}
\int_\zeta \hspace{-0.mm}
p(\xi) p(\eta) p(\zeta) \
\delta \left( \Delta u'_r\hspace{-0.mm}-\hspace{-0.mm}\Delta {u}_r(\xi, \eta ,\zeta) \right)   
d \xi \ d \eta \ d \zeta
\end{array}
\label{frobenious_perron}
\eea 
where $\Delta {u}_r$ is calculated with Eq. (\ref{fluc4}), $\delta$ is the Dirac delta and $p$ is a gaussian PDF whose average value and standard deviation are equal to 0 and 1, respectively.

\bigskip

For non-isotropic turbulence or in more complex cases
with boundary conditions, the velocity spectrum could not satisfy the
Lindeberg condition, thus the velocity will be not distrubuted following 
a Gaussian PDF, and Eq. (\ref{fluc3}) changes its analytical form and can incorporate more intermittant terms \cite{Lehmann99} which give the deviation with respect to the isotropic turbulence.
Hence, the absolute statistical moments of $\Delta {u}_r$ will be greater than those calculated with Eq. (\ref{fluc4}), indicating that, in a more complex situation than the isotropic turbulence, the intermittency of $\Delta {u}_r$ can be significantly stronger.

\section{\bf Results and discussion \label{s8}}

In order to validate the results of the proposed Lyapunov analysis, several data 
are now presented.

As the first result, the evolution in the time of $f$ is calculated with the proposed closure (Eq. (\ref{vk6})) of the von K\'arm\'an-Howarth equation, where the boundary conditions are given by Eq. (\ref{bc}) (see Appendix).
The turbulent kinetic energy and $E(\kappa)$ and $T(\kappa)$ are calculated with Eq. (\ref{ke}) and Eqs. (\ref{Ek}), respectively.
The calculation is carried out for an initial Reynolds number of 
$Re(0) = u(0) L_r/ \nu$ = 2000, where $L_r$ and $u(0)$ are, respectively, 
the reference dimension and the initial velocity standard deviation. 
The initial condition for $f$ is 
$f(r) = \exp\left( -1/2 (r/\lambda_T)^2\right) $, where 
$\lambda_T/L_r$ = $1/(2 \sqrt{2})$, whereas $u(0)$ = 1. 
The dimensionless time of the problem is defined as $\bar{t} = t \ u(0)/L_r$.
\begin{figure}
\vspace{-0.mm}
	\centering
\includegraphics[width=0.45\textwidth]{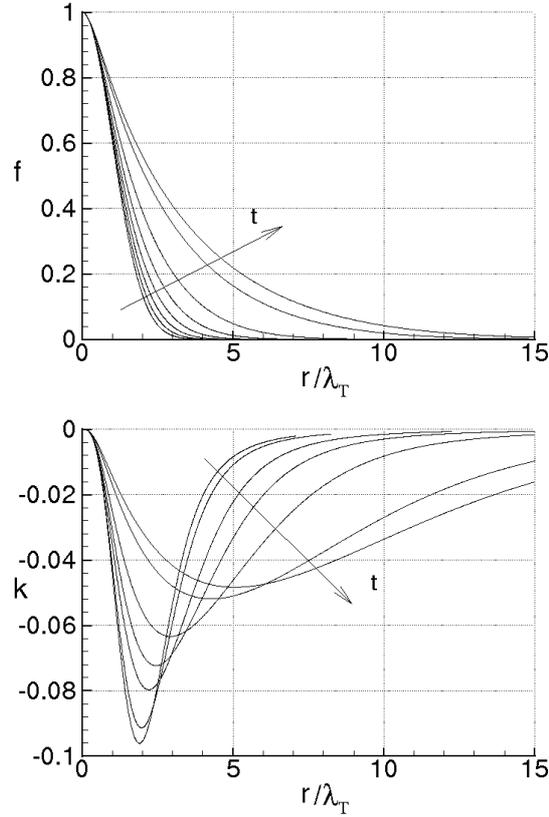}
\vspace{-0.mm}
\caption{Correlation functions, $f$ and $k$ versus the separation distance at the times of simulation $\bar{t}$ = 0, 0.1, 0.2, 0.3, 0.4, 0.5, 0.6, 0.63, $Re(0) = 2000$.}
\vspace{-0.mm}
\label{figura_6}
\end{figure}

Equation (\ref{vk}) was numerically solved adopting the Crank-Nicholson integrator scheme with variable time step $\Delta t$, whereas the discretization in the spatial domain is made by $N-1$ intervals of the same amplitude $\Delta r$.
The truncation error of the scheme of integration is 
${\cal O}(\Delta t^2) + {\cal O} (\Delta r^2)$, where $\Delta t$ is automatically 
selected by the algorithm in such a way that
${\cal O}(\Delta t^2) \sim {\cal O} (\Delta r^2)$ for each time step, thus
the accuracy of the scheme is of the order of ${\cal O} (\Delta r^2)$.
As the consequence, the discretization of the Fourier space is made by $N-1$ subsets in the interval $\left[ 0, \kappa_M \right]$, where $\kappa_M$ = $\pi/(2 \Delta r)$.
For the adopted initial Reynolds number the choice $N$ = 1500 gives an adequate discretization, which provides $\Delta r < \ell$, for the whole simulation.
For what concerns $u$, it was calculated with Eq. (\ref{ke}) and the kinetic energy was checked to be equal to the integral over $\kappa$ of $E(\kappa)$. 
During the simulation, $T(\kappa)$ must identically satisfy Eq.(\ref{tk0})
(see Appendix) which states that $T(\kappa)$ does not modify the kinetic energy.
The integral of $T(\kappa)$ is calculated with the trapezoidal formula, 
for $\kappa \in (0, \kappa_M)$, and the simulation will be considered to be accurate
 as long as
\bea
\int_0^{\kappa_M} T(\kappa) d \kappa \simeq \int_0^{\infty}
 T(\kappa) d \kappa = 0
\label{tk0a}
\eea   
namely, when $T(\kappa) \simeq 0$ for $\kappa > \kappa_M$.
As the simulation advances, according to Eq. (\ref{vk6}), the energy cascade
determines variations of $E(\kappa)$ and $T(\kappa)$ for values of $\kappa$ which 
rise with the time and that can occur out of the interval ($0$, $\kappa_M$).
Thus, Eq. (\ref{tk0a}) holds until a certain time, where these wave-numbers
 are about equal to $\kappa_M$.
For higher times, the variations of $T(\kappa)$ occur for $\kappa > \kappa_M$, 
out of the interval of numerical integration $(0,  \kappa_M)$, 
and Eq. (\ref{tk0a}) could not be satisfied.
Hence, the results of the simulation will be accurate until reaching of the
following condition \cite{NAG}
\bea
\ds \left| \int_0^{\kappa_M} T(\kappa) d \kappa \right| \geq \frac{1}{12} \frac{\kappa_M^3}{N^2} 
\left| \frac{\partial^2 T}{\partial \kappa^2} \right|_{max} \approx
 \frac{1}{N^2} \int_0^{\kappa_M} \vert T(\kappa) \vert  d \kappa
\label{accur}
\eea
where the right-hand side represents the estimation of the truncation error of
the trapezoid formula \cite{NAG}.
There, it is found that, $\Delta r \approx 0.8 \ \ell$ in all the simulations. .
Thereafter, the numerical method loses its consistency, 
since the violation of Eq. (\ref{tk0a}) implies that $K$ does not preserve 
the kinetic energy. 
Nevertheless, the error committed by the algorithm is considered still acceptable as long as 
\cite{Karman38}, \cite{Batchelor53}
\bea
\ds 0 <
 \left| \int_0^{\kappa_M} T(\kappa) d \kappa \right|
 <<< \left|  \frac{d u^2}{dt}  \right| 
\label{accur1}
\eea
That is to say, when the numerical residual of the rate of energy transfer is much less than
the dissipation rate. In any case the simulations are stopped if $\ell$ reaches its relative minimum in function of the time with $\ell > \Delta r$, or as soon as $\Delta r = \ell$.
The accuracy of the solutions is evaluated through the check of Eq. (\ref{accur1}).
At end of simulation, we assume that the energy spectrum is fully developed.

\bigskip

In order to study the proposed closure, we first analyze the case with $K=0$.
In this case, with the assumed initial condition, Eq. (\ref{vk}) admits 
the analytical self-similar solution \cite{Karman38}
\bea
f(t, r) = \exp\left( -\frac{1}{2}  \left( \frac{r}{\lambda_T(t)}\right) ^2\right), \ \mbox{where} \ 
 \lambda_T(t)=\sqrt{\lambda^2_T(0) + 4 \nu t}
\label{s_s}
\eea 
$f(t, r)$ maintains its shape unchanged in function of the dimensionless coordinate
 $r/\lambda_T(t)$. 
The corresponding numerical solutions were calculated 
for different spatial discretization (i.e. $N$ = $500$, $1000$, and  $1500$).
All these simulations satisfy Eq. (\ref{s_s}), with a calculated 
error which is always of the order of the truncation error of the scheme of integration.
Also the energy spectrum preserves its shape, and is given by
$\ds E(\kappa  \lambda_T) = A u^2 (\kappa \lambda_T)^4 \exp(-a (\kappa \lambda_T)^2)$, 
with $A =O(1)$,  $a=O(1)$ \cite{Batchelor53}.

\bigskip

Consider now the case, where $K$ is given by Eq. (\ref{vk6}).
The diagrams of Fig. \ref{figura_6} show the evolution of $f(r)$ and $k(r)$ in terms of  $r/\lambda_T$, at different times of simulation.
The kinetic energy and $\lambda_T$ vary according to Eqs. (\ref{vk6}),  (\ref{ke})
and (\ref{dl/dt}), thus $f(r)$ and $k(r)$ change in such a way that their scales, 
in particular $\ell$ and $\lambda_T$, diminish as the time increases, whereas the maximum of 
$\vert k \vert$ decreases. 
That is, Eq. (\ref{vk6}) corresponds to a transferring of the energy toward the 
smaller scales which strongly contrasts the effects of viscosity seen in the previous case.

At the final instants of the simulation, one obtains that $f - 1 = $ O( $r^{2/3}$) 
for $r/\lambda_T=$ O(1), and  the maximum of $\vert k \vert$ is about 0.05.
These results are in very good agreement with the numerous data of the literature \cite{Batchelor53} which concern the evolution of $f$ in homogeneous isotropic turbulence.
\begin{figure}
\vspace{-0.mm}
	\centering
\hspace{-0.mm}
\includegraphics[width=0.49\textwidth]{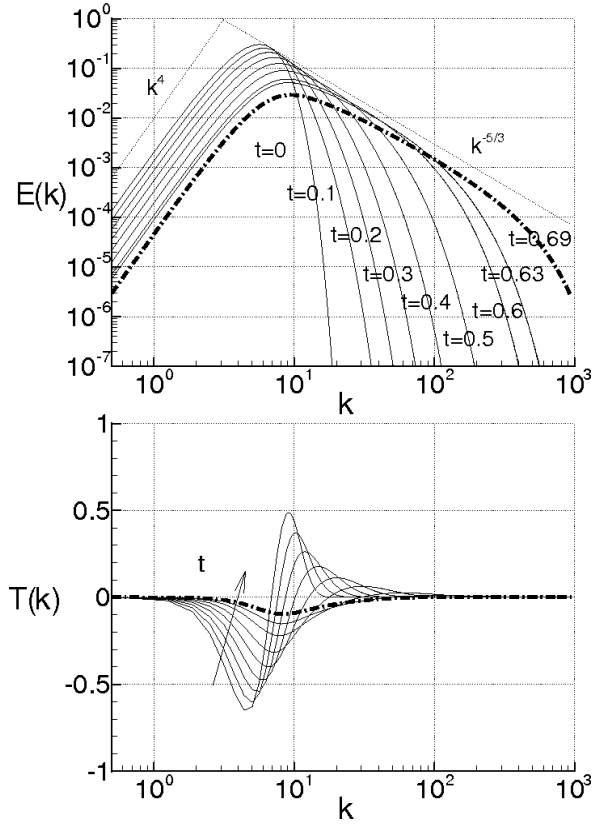}
\vspace{-0.mm}
\caption{Plot of $E(\kappa)$ and $T(\kappa)$, for $Re(0)=2000$, at the diverse times of simulation.}
\vspace{-0.mm}
\label{figura_7}
\end{figure}
Figure \ref{figura_7} shows the diagrams of $E(\kappa)$ and $T(\kappa)$, at the same times, in comparison with the Kolmogorov law ($\kappa^{-5/3}$) and with the incompressibility condition ($\kappa^4$).
The energy spectrum depends on the initial condition, and at the end of the simulation, 
can be compared with the Kolmogorov spectrum in an opportune interval of wave-numbers which defines the inertial subrange.
This arises from $f$, which, at the final times, behaves like $f -1$ = O ($r^{2/3}$)
 for $r = O(\lambda_T)$. 
$E(\kappa)$ satisfies the continuity equation which implies that 
$E(\kappa) \approx \kappa^4$ near the origin.
\begin{figure}
\vspace{-0.mm}
	\centering
\hspace{-0.mm}
\includegraphics[width=0.45\textwidth]{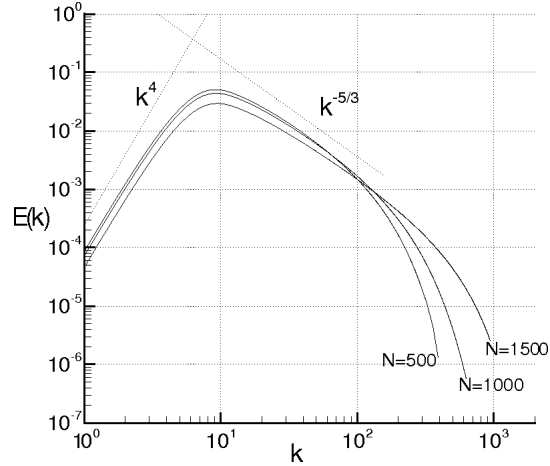}
\vspace{-0.mm}
\caption{Plot of the final energy spectrum for different spatial discretization,
$Re(0)=2000$.}
\vspace{-0.mm}
\label{figura_7A}
\end{figure}
In the figure, the dimensionless time $\bar{t} = 0.63$ corresponds to the condition (\ref{accur}), whereas $\bar{t} = 0.69$ (bold dashdotted line) represents the spectrum calculated when $\ell$ reaches its minimum in function of the time.
In this last situation, the algorithm is less accurate, due to the 
discretization. In fact, $T(\kappa)$, (bold dashdotted), exhibits small nonzero values in a wide range of variations of $\kappa$, for $\kappa > \kappa_M$.
Although Eq. (\ref{tk0a}) is not satisfied, the error of the algorithm is
still acceptable since the absolute value of the integral of $T(\kappa)$ 
over ($0, \kappa_M$) is about four orders of magnitude less than the dissipation rate 
(see Eq. (\ref{accur1})).
\begin{figure}
\vspace{-0.mm}
	\centering
\hspace{+10.mm}
\includegraphics[width=0.65\textwidth]{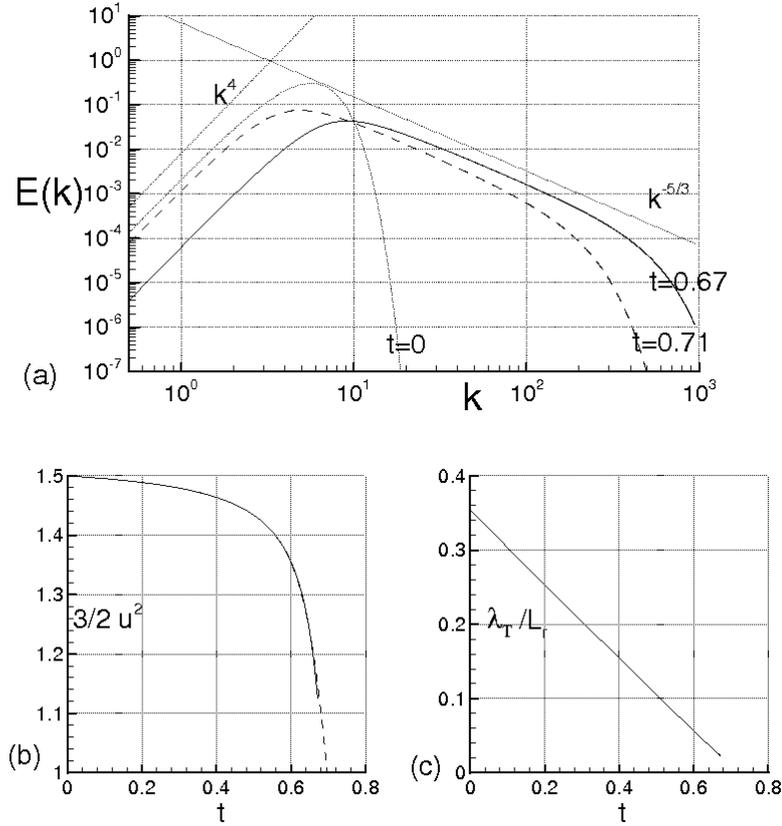}
\vspace{-0.mm}
\caption{Comparison of the results for Re(0)=3000: Continuous line: present analysis.
Dashed line:  Oberlack's model (Ref. \cite{Oberlack93}). (a) Dotted curve: initial condition.
(b) Kinetic energy, (c) Taylor scale}
\vspace{-0.mm}
\label{figura_9_A}
\end{figure}


To study the effect of the spatial discretization on the final energy spectrum,
Fig. \ref{figura_7A} shows the results of other simulations which
were performed for $N$ = 500, 1000.
It is found that, for $N$ = 500 and 1000, the stopping criterion is satisfied for 
$\ell = \Delta r$, whereas the final times of simulation rises with $N$, resulting that 
$\bar{t} \approx 0.33$ and $\bar{t} \approx 0.64$ for $N$ = 500 and 1000, respectively.
As the consequence, also the inertial subrange of Kolmogorov increase with $N$.

\bigskip

In order to compare the results of the present analysis with the data in the literature,
a further calculation has been carried out with an initial Reynolds number of $Re(0) = 3000$.
The present results (continuous line) are 
shown in Fig. \ref{figura_9_A}, in terms of
energy spectrum, and are
compared with those calculated, for the same conditions,
with the Oberlack's model \cite{Oberlack93} (dashed line).
For this latter, the parameter $k_2$ \cite{Oberlack93} (see introduction) is 
assumed to be equal to $k_2 = -H_3(0)/(\sqrt{2} \ 6)$. For this value of $k_2$, the Oberlack's model gives the same value of skewness $H_3(0) = -3/7$, calculated with the present analysis. Thus, the entity of the mechanism of the energy cascade is the same in both the cases.
Again, the initial correlation function is $f = \exp(-(r/\lambda_T)^2/2)$, 
$\lambda_T/L_r = 1/(2 \sqrt{2})$ and $N=$ 1500.

Equation  (\ref{tk0a}) is satisfied until a certain time 
(i.e.  $t \approx 0.66$ for Oberlack's model and $t \approx 0.62$ for the present analysis), therafter the numerical scheme exhibits a minor accuracy, since
$T(\kappa) \ne 0$ for $\kappa > \kappa_M$). 
In the figure, the energy spectrum is shown at the end of the two simulations which
happen at $\bar{t} = 0.71$ (Oberlack) and at $\bar {t} = 0.67$ (Present Analysis),
where the stopping condition is satisfied for  $\ell=$ min $\approx \Delta r$.
There, the integral of $T(\kappa)$ is, in both the cases, much less than $\vert d u^2/ dt \vert$ (about four orders of magnitude) and for this reason the results of the two simulations can be considered accurate enough.
Since the skewness $H_3(0)$ is the same in the two cases, the variations of 
$\lambda_T$ almost coincide during the simulations, a part very small variations caused by $\nu$ (see Eq. (\ref{dl/dt})) which are not appreciated in the figure.
Therefore, also $d u^2/dt$ is about the same in both the cases.
Nevertheless, the energy spectra show significative differences.
The spectrum calculated with 
Eq. (\ref{vk6}) exhibits a wider range of wave-numbers than the other one, and this is due to the fact that Eq. (\ref{vk6}) represents a nonlinear term of the first order
which does not cause any diffusion effect, whereas for the Oberlack's model, the closure
term is of the second order and the variable eddy diffusivity produces a sizable reduction of all the wave-numbers, especially for large $\kappa$. 
The wave-number calculated where $E(\kappa) = 10^{-6}$ (see figure), is quite different in the two cases.
Specifically, the present analysis gives a value two times greater than in the
Oberlack's model, and, as the consequence, also the Kolmogorov inertial subrange calculated 
with the present analysis is about two times the other one.

\bigskip

Next, the Kolmogorov function $Q(r)$ and Kolmogorov constant $C$, are determined, 
using the results of the previous simulations. 

\begin{figure}
\vspace{+0.mm}
	\centering
\hspace{-0.mm}
\includegraphics[width=0.42\textwidth]{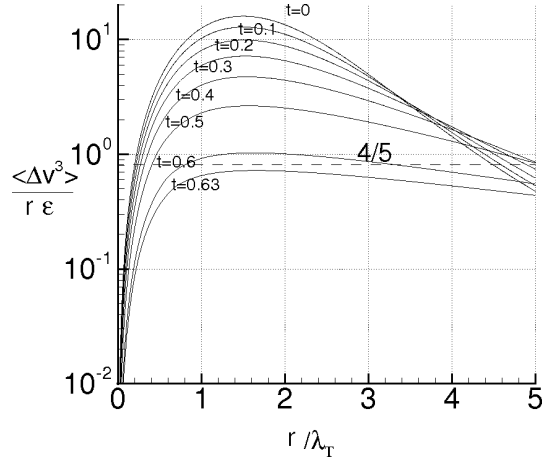}
\vspace{-0.mm}
\caption{The Kolmogorov function versus $r/\lambda_T$. at $Re(0)=2000$, for different times of simulation. The dashed line indicates  the value  4/5.}
\vspace{-0.mm}
\label{figura_8}
\end{figure}

Following the Kolmogorov theory, the Kolmogorov function, which is defined as  
\bea 
\ds Q(r) = - \frac{\langle (\Delta u_r)^3 \rangle} { r \varepsilon}
\label{k_f}
\eea
is constant w. r. t. $r$, and is equal to 4/5 as long as $r/\lambda_T = O(1)$. 
As shown in Fig. \ref{figura_8}, for $\bar{t} =0$, $Q_{max}$ is significantly 
greater than $4/5$ and the variations of $Q$ with $r/\lambda_T$ can not be neglected.
This is the consequence of the choice of the initial correlation function.
At the successive times, $Q_{max}$ decreases until the final instants, where, with the exception of $r/\lambda_T \approx 0$, $Q(r)$ exhibits variations which are less than those calculated at the previous times in a wide range of $r/\lambda_T$, with a maximum which can be compared to 0.8.
The Kolmogorov constant $C$ is also calculated by 
\bea
C = \max_{\kappa \in (0, \kappa_M)} 
\left( \frac{E(\kappa) \kappa^{5/3}} {\varepsilon^{2/3}} \right)
\label{k_c}
\eea 
For $Re(0) = 2000$ and $\bar{t}\simeq$ 0.63, $C \simeq  1.932$ and $Q_{max} \simeq .73$,
whereas for $\bar{t}\simeq$ 0.69, $C \simeq  1.92$, $Q_{max} \simeq .72$.
For $Re(0) = 3000$ at end simulation (i.e. $\bar{t}\simeq$ 0.67), 
$C \simeq 1.941$, $Q_{max} \simeq .75$, 
namely $C$ and $Q_{max}$ agree with the corresponding quantities known from the literature.

\begin{figure}
\vspace{-0.mm}
	\centering
\hspace{+10.mm}
\includegraphics[width=0.45\textwidth]{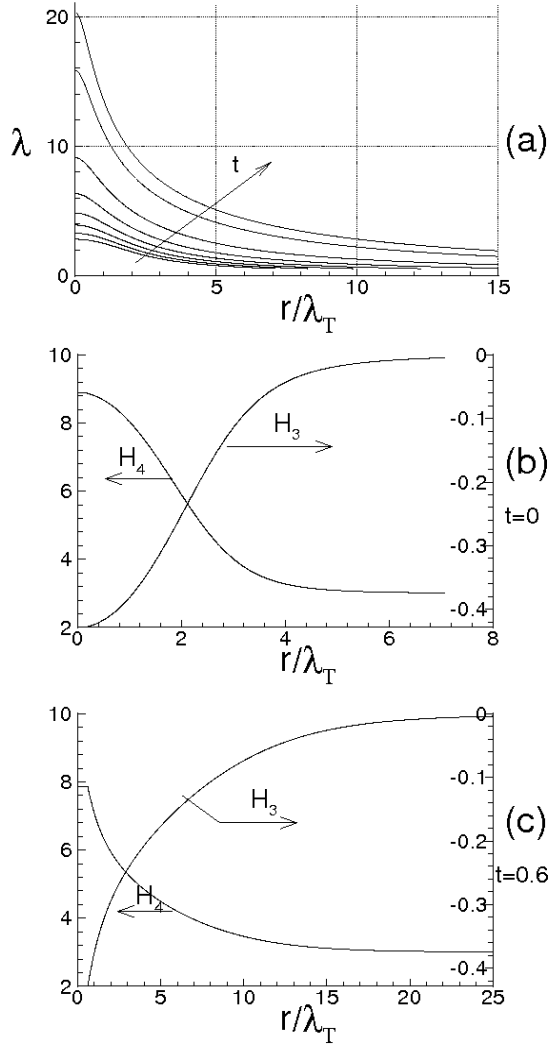}
\vspace{-0.mm}
\caption{(a) Maximum finite size Lyapunov exponent calculated at $Re(0)=2000$, at the times of simulation $\bar{t}$ = 0, 0.1, 0.2, 0.3, 0.4, 0.5, 0.6, 0.63; (b) and (c) skewness and Flatness versus $r/\lambda_T$ at t = 0 and t = 0.6, respectively.}
\vspace{-0.mm}
\label{figura_9}
\end{figure}

For $Re(0) = 2000$, Fig. \ref{figura_9}a shows the maximal finite scale Lyapunov exponent, calculated with Eq. (\ref{lC}).
For $t = 0$, the variations of $\lambda$ are the result of the adopted initial
correlation function which is a gaussian, whereas as $t$ increases, the
variations of $f$ determine sizable increments of $\lambda$ and of its slope in
proximity of the origin.
Then, for $t = 0.6$, since $f -1 \approx O (r^{2/3})$, 
the maximal finite scale Lyapunov exponent behaves like $\lambda \approx r^{-2/3}$.
Thus, the Richardson's diffusivity associated to the relative motion between
two fluid particles,  defined as $D_R \propto \lambda r^2$ \cite{Richardson26}, here
satisfies the famous Richardson scaling law $D_R \approx  r^{4/3}$\cite{Richardson26}
in proximity of the end of the simulation.

\bigskip

\begin{figure}
\vspace{-0.mm}
	\centering
\includegraphics[width=0.48\textwidth]{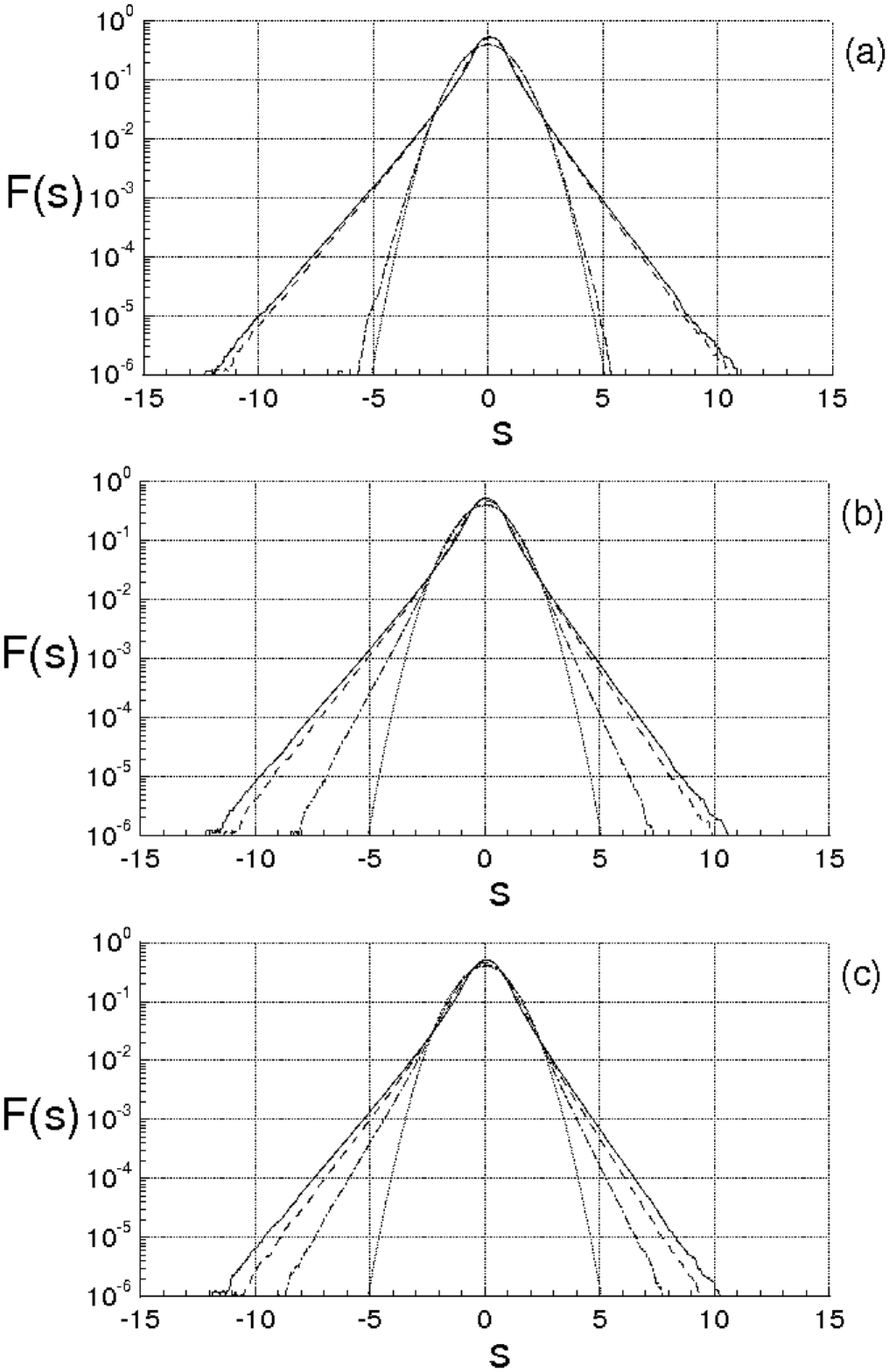}
\vspace{-0.mm}
\caption{PDF of the velocity difference fluctuations at the times $\bar{t}=$0 (a), $\bar{t}=$ 0.5 (b) and $\bar{t}=$0.6 (c). 
Continuous lines are for $r=$0, dashed lines are for $r/\lambda_T =$1, 
dot-dashed lines are for  $r/\lambda_T =$5, dotted lines are for gaussian PDF.}
\vspace{-0.mm}
\label{figura_10}
\end{figure}

In the figures \ref{figura_9}b and \ref{figura_9}c, skewness and flatness of $\Delta u_r$ are shown in terms of $r$ for $\bar{t}$ = 0 and 0.6 and $Re(0) = 2000$.
The skewness, $H_3$ is first calculated with Eq. (\ref{H_3_01}), 
then $H_4$ has been determined using Eq. (\ref{m1}). 
At $\bar{t}=0$, $\vert H_3 \vert$ starts from 3/7 at the origin with small slope, then decreases until reaching small values. $H_4$ also exhibits small derivatives near the origin, where $H_4\gg$ 3, thereafter it decreases more rapidly than $\vert H_3 \vert$. 
At $\bar{t} = $0.6, the diagram importantly changes and exhibits different shapes.
The Taylor scale and $R_\lambda$ are both changed, so that the variations of 
$H_3$ and $H_4$ are associated to smaller distances, whereas the flatness at the origin is slightly less than that at $t =0$.
Nevertheless, these variations correspond to higher values of  $r/\lambda_T$ than those for $t$ = 0, 
and again $H_4$ reaches the value of 3 more rapidly than $H_3$ tends to zero.

The PDFs of $\Delta u_r$ are calculated, for $Re(0)=2000$, with Eqs. (\ref{frobenious_perron})
and (\ref{fluc4}), and are shown in Fig. \ref{figura_10} in terms of the dimensionless abscissa 
\bea
\ds s = \frac{\Delta u_r} 
{ \langle (\Delta u_r)^2 \rangle^{1/2}  }
\nonumber
\eea
where, these distribution functions are normalized, in order that their standard 
deviations are equal to the unity.
The figure  represents the distribution functions of $s$ for several
$r/\lambda_T$, at $\bar{t}$ = 0, 0.5 and 0.6, where the dotted curves represent the gaussian distribution functions.
The calculation of $H_3(r)$ is first carried out with Eq. (\ref{H_3_01}), 
then $\psi(r, R_\lambda)$ is identified through Eq. (\ref{H_3}), 
and finally the PDF is obtained with  Eq. (\ref{frobenious_perron}).
For $t$ = 0 (see Fig. \ref{figura_10}a) and according to the evolutions of
 $H_3$ and $H_4$, the PDFs calculated at $r/\lambda_T=$ 0 and 1, are quite similar each 
other, whereas for $r/\lambda_T=$ 5, the PDF is almost a gaussian function.
Toward the end of the simulation, (see Fig. \ref{figura_10}b and c), the two PDFs
calculated at $r/\lambda_T=$ 0 and 1, exhibit more sizable differences, whereas for $r/\lambda_T=$ 5, the PDF differs very much from a gaussian PDF. 
This is in line with the plots of $H_3(r)$ and $H_4(r)$ of Fig. \ref{figura_9}.
\begin{figure}
\vspace{-0.mm}
	\centering
\vspace{+4.mm}
\includegraphics[width=0.45\textwidth]{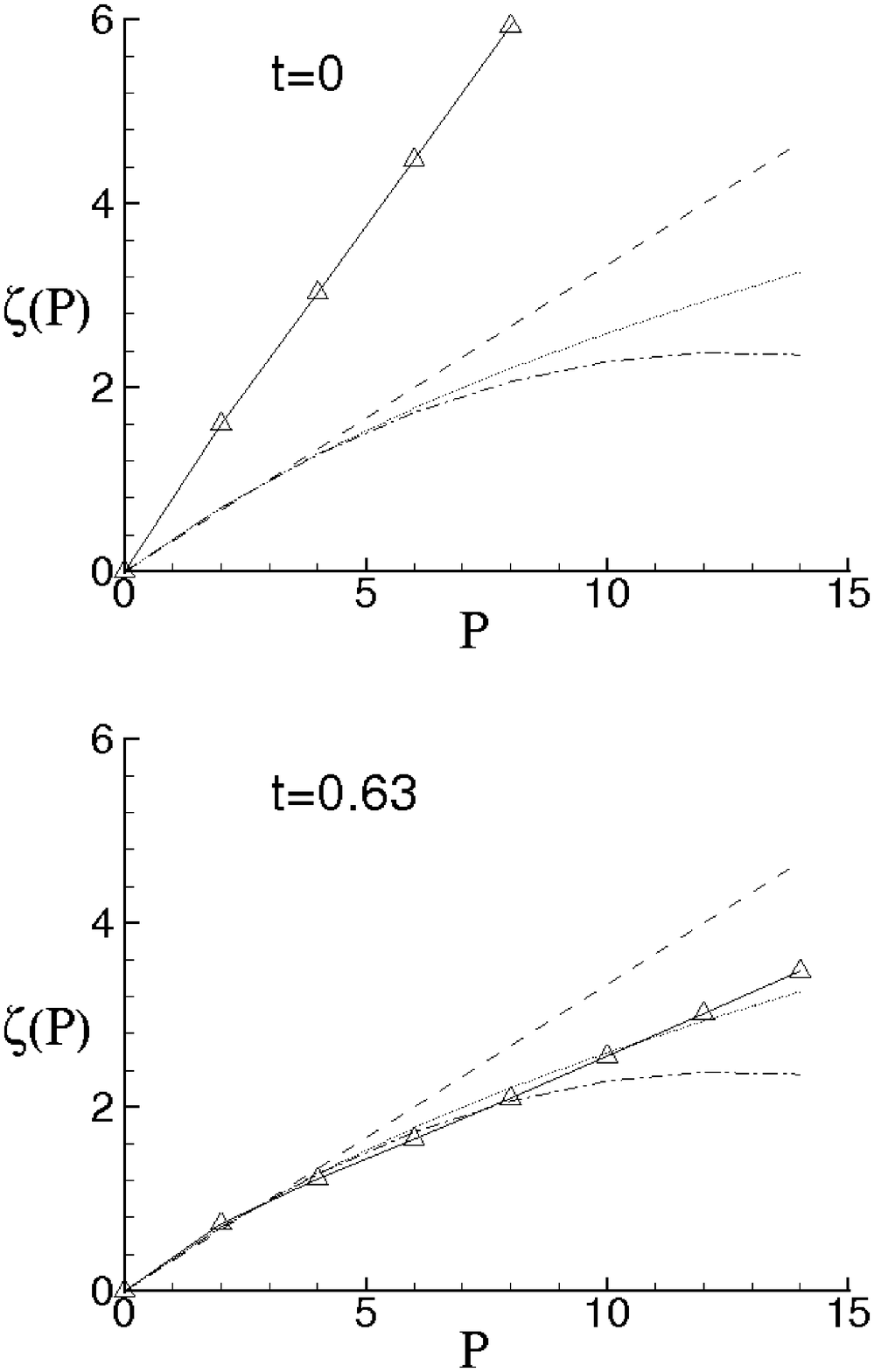}
\vspace{-0.mm}
\caption{Scaling exponents of the longitudinal velocity difference versus the order moment at different times, $Re(0)=2000$. Continuous lines with solid symbols are for the present data. Dashed lines are for Kolmogorov K41 data \cite{Kolmogorov41}. Dashdotted lines are for Kolmogorov K62 data \cite{Kolmogorov62}.
 Dotted lines are for She-Leveque data \cite{She-Leveque94}}
\vspace{-0.mm}
\label{figura_11}
\end{figure}

\bigskip

Next, the spatial structure of $\Delta u_r$, given by Eq. (\ref{fluc4}), is analyzed 
using the previous results. 
According to the various works \cite{Kolmogorov62}, \cite{She-Leveque94}, 
$\Delta u_r$ behaves quite similarly to a multifractal system, where $\Delta u_r$ 
obeys to a law of the kind 
$
\Delta u_r(r) \approx r^q
$
where the exponent $q$ is a fluctuating function of the space coordinates.
This implies that the statistical moments of $\Delta u_r(r)$ are expressed through 
different scaling exponents $\zeta(P)$ whose values depend on the moment order $P$, i.e.
\bea
\left\langle (\Delta u_r)^{P}(r) \right\rangle  = A_P \ r^{\zeta(P)}
\label{fractal}
\eea
\begin{table}[b] 
\caption{Scaling exponents of the longitudinal velocity difference for different 
initial Reynolds numbers.}
  \begin{center} 
  \begin{tabular}{lrrrrrrrrrrrrrrr}
\vspace{0.mm}
Re(0) & P       & 1   & 2     & 3     & 4    & 5    & 6    & 7    & 8    & 9   & 10  & 11   & 12   & 13   & 14   \\
\hline
\hline \\
2000 & $\zeta$(P) & 0.36 & 0.71 & 1.00 & 1.19 & 1.41 & 1.61 & 1.84 & 2.04 & 2.25 & 2.49 & 2.72 & 2.93 & 3.15 & 3.38 \\
3000 & $\zeta$(P) & 0.36 & 0.72 & 1.00 & 1.19 & 1.43 & 1.64 & 1.84 & 2.03 & 2.25 & 2.49 & 2.71 & 2.92 & 3.11 & 3.33 \\
\hline
 \end{tabular}
  \end{center} 
\end{table} 
These scaling exponents are here identified through a best fitting procedure, 
in the interval ($a, a$ $ + \lambda_T$), where the endpoints $a$ is an unknown 
quantity which has  to be determined.
The location of this interval varies with the time. 
The calculation of $a$,  $\zeta_P$ and $A_P$ is carried out through a minimum square method 
which, for each moment order, is applied to the following optimization problem
\bea
\ds J_P(\zeta_P, A_P) \hspace{-1.mm} \equiv 
\int_{a}^{a +\lambda_T} 
\ds ( \langle (\Delta u_r)^P \rangle - A_P r^{\zeta(P)} )^2 dr  = \mbox{min}, \  
 P = 1, 2, ...
\eea
where $(\langle \Delta u_r^P)\rangle$ are calculated with Eqs. (\ref{m1}), and $a$ is calculated 
in order to obtain $\zeta (3) = 1$.
\\
Figure \ref{figura_11} shows the comparison between the scaling exponents here obtained  (continuous lines with solid symbols) and those of the Kolmogorov theories K41 \cite{Kolmogorov41} (dashed lines) and K62 \cite{Kolmogorov62} (dashdotted lines), and those given by She-Leveque \cite{She-Leveque94} (dotted curves).
At $t=$ 0, the values of $\zeta(P)$ are the result of the chosen initial condition.
As the time increases, $f$ changes causing variations in the statistical
moments of $\Delta u_r(r)$. As result, $\zeta(P)$ gradually diminish and exhibit a variable slope which depends on the moment order $P$, until to reach the situation of 
Fig. \ref{figura_11}b, where the simulation is just ended. 
The dimensionless moments of $\Delta u_r(r)$ are changed. The plot of $\zeta(P)$ shows that near the origin, $\zeta(P) \simeq P/3$, and that the values of $\zeta(P)$ seem to be in agreement with the those proposed by She-Leveque.
More in detail, Table I reports these scaling exponents in terms of the moments order, calculated for $\bar{t} = 0.63$.
These values and the peculiar diagram of Fig. (\ref{figura_11}) are the consequence of the spatial variations of the skewness, calculated using Eq. (\ref{H_3_01}), and of the quadratic terms due to the inertia and pressure forces into the expression of the velocity difference, which make 
$\langle (\Delta u_r)^P \rangle$ a quantity quite similar to a multifractal system.

Other simulations with different initial correlation functions and Reynolds
numbers have been carried out, and all of them lead to analogous results,
in the sense that, at the end of the simulations, the diverse quantities 
such as $Q(r)$, $C$ and $\zeta(P)$ are quite similar to those just calculated.
For what concerns the effect of the Reynolds number, its increment determines
a wider range of the wave-numbers where $E(\kappa)$ is comparable with the Kolmogorov 
law and a smaller dissipation energy rate in accordance to Eq. (\ref{ke}).
\bigskip
\begin{table}[b] 
\caption{Dimensionless statistical moments of $F(\partial u_r/\partial r)$ at different
Taylor scale Reynolds numbers. P.R. as for ''present results''.}
  \begin{center} 
  \begin{tabular}{lrrrr} 
\hline
Moment \ & $R_\lambda \approx 10$ \ & $R_\lambda=10^2$ \ & $R_\lambda=10^3$ \ & Gaussian\\[2pt] 
Order    &  P. R.                 & P. R.            &  P. R. \           & Moment \\[2pt] 
\hline \\
3        & -.428571               & -.428571         & -.428571          & 0      \\
4        &   3.96973              &  7.69530         & 8.95525           & 3      \\
5        & -7.21043               &  -11.7922        & -12.7656          & 0      \\
6        &  42.4092               &  173.992         & 228.486           & 15     \\
7        & -170.850               &  -551.972        & -667.237          & 0      \\
8        &  1035.22               &  7968.33         & 11648.2           & 105    \\
9        &  -6329.64              &  -41477.9        & -56151.4          & 0      \\
10       & 45632.5                &  617583.         & 997938.           & 945    \\
\hline
 \end{tabular}
  \end{center} 
  \vspace{-5.mm}
\end{table} 

\begin{figure}[t]
\vspace{-0.mm}
	\centering
\includegraphics[width=0.45\textwidth]{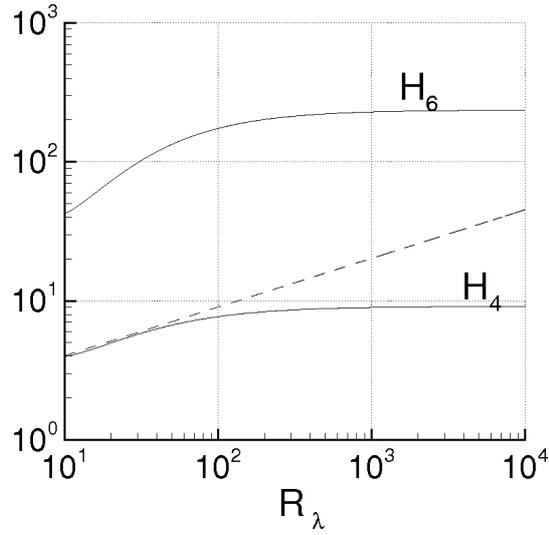}
\vspace{-0.mm}
\caption{Dimensionless moments $H_4(0)$ and $H_6(0)$ plotted vs. $R_{\lambda}$.
Continuous lines are for the present results. 
The dashed line is the tangent to the curve of $H_4(0)$ in $R_\lambda \approx$ 10.}
\vspace{-0.mm}
\label{figura_3}
\end{figure}

In order to study the evolution of the intermittency vs. the Reynolds number,
Table II gives the first ten statistical moments of $F(\partial u_r/\partial r)$. 
These are calculated with Eqs. (\ref{m1}) and (\ref{m2}), for $R_{\lambda}$ = 10.12, 100 and 1000, and are shown in comparison with those of a gaussian distribution function.
It is apparent that a constant nonzero skewness of $\partial u_r /\partial r$, causes an intermittency which rises with $R_\lambda$ (see Eq. (\ref{fluc4})).
More specifically, Fig. \ref{figura_3} shows the variations of $H_4(0)$ and $H_6(0)$
(continuous lines) in terms of $R_\lambda$, calculated with Eqs. (\ref{m1}) 
and (\ref{m2}), with $H_3(0) = -3/7$.
These moments are rising functions of $R_{\lambda}$ for 
10 $< R_{\lambda} <$ 700, whereas for higher $R_{\lambda}$ these tend to
the saturation and such behavior also happens for the other absolute moments.
According to Eq. (\ref{m1}), in the interval 10 $ < R_{\lambda} <$  70, 
$H_4$ and $H_6$ result to be about proportional to $R_{\lambda}^{0.34}$ and $R_{\lambda}^{0.78}$, respectively, and the intermittency increases with the
 Reynolds number until $R_{\lambda} \approx$ 700,  
where it ceases to rise so quickly. 
\begin{figure}[t]
\vspace{-0.mm}
	\centering
\includegraphics[width=0.48\textwidth]{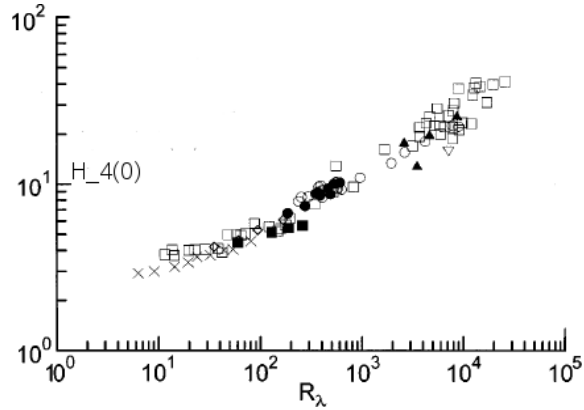}
\vspace{-0.mm}
\caption{Flatness $H_4(0)$ vs. $R_{\lambda}$. These data are from  Ref.\cite{Antonia97}.}
\vspace{-0.mm}
\label{antonia}
\end{figure}
\begin{figure}[t]
\vspace{-0.mm}
	\centering
\includegraphics[width=0.50\textwidth]{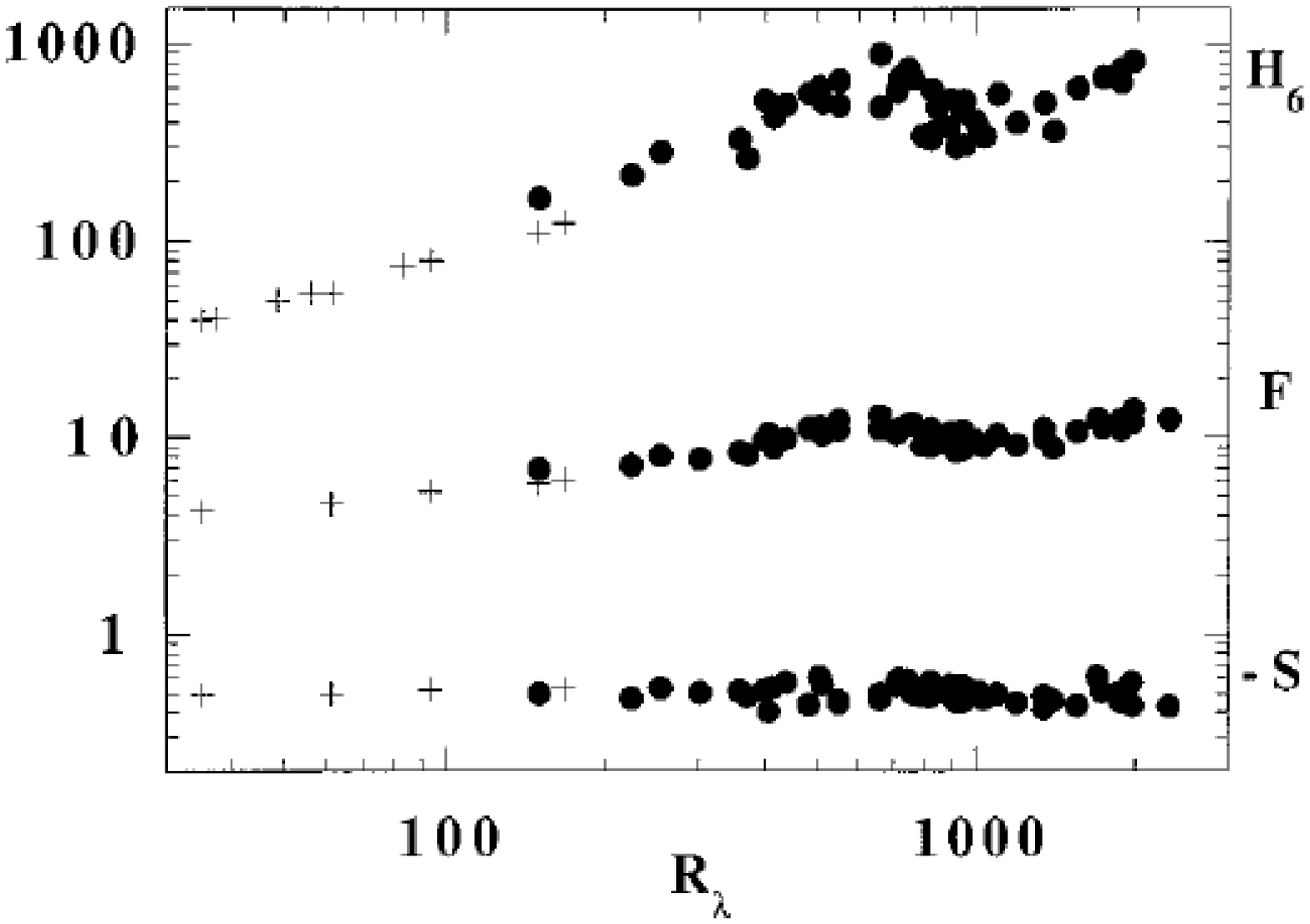}
\vspace{-0.mm}
\caption{Skewness $S = H_3(0)$, Flatness $F =H_4(0)$ and hyperflatness $H_6(0)$ vs. $R_{\lambda}$. These data are from  Ref.\cite{Tabeling97}.}
\vspace{-0.mm}
\label{tabeling0}
\end{figure}
This behavior, represented by the continuous lines, depends on the fact that 
$\psi \approx \sqrt{R_\lambda}$, and results to be in very good agreement with the data 
of Pullin and Saffman \cite{Pullin93}, for 10 $< R_{\lambda} <$  100.
\begin{figure}[t]
\vspace{+0.mm}
\centering
\hspace{0.mm}
\includegraphics[width=0.45\textwidth]{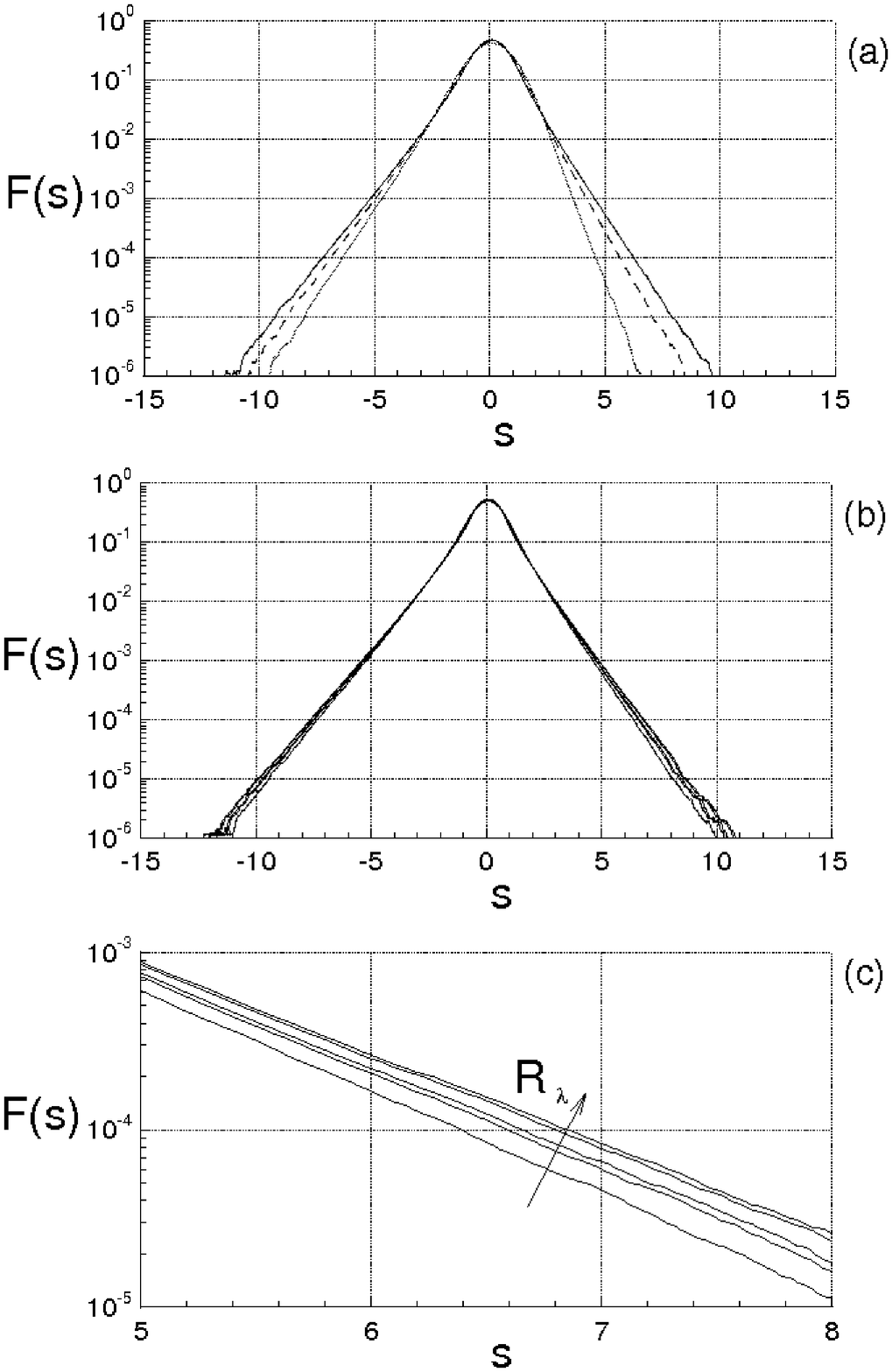}
\vspace{-0.mm}
\caption{Log linear plot of the PDF of $\partial u_r/\partial r$ for different
$R_\lambda$. (a): dotted, dashdotted and continuous lines are
for $R_\lambda$ = 15, 30 and 60, respectively. (b) and (c) PDFs for 
$R_{\lambda}$ = 255, 416, 514, 1035 and 1553. (c) represents an enlarged part
of the diagram (b) }
\vspace{-0.mm}
\label{figura_4}
\end{figure}
\begin{figure}[t]
\vspace{+4.mm}
\centering
\hspace{0.mm}
\includegraphics[width=0.42\textwidth]{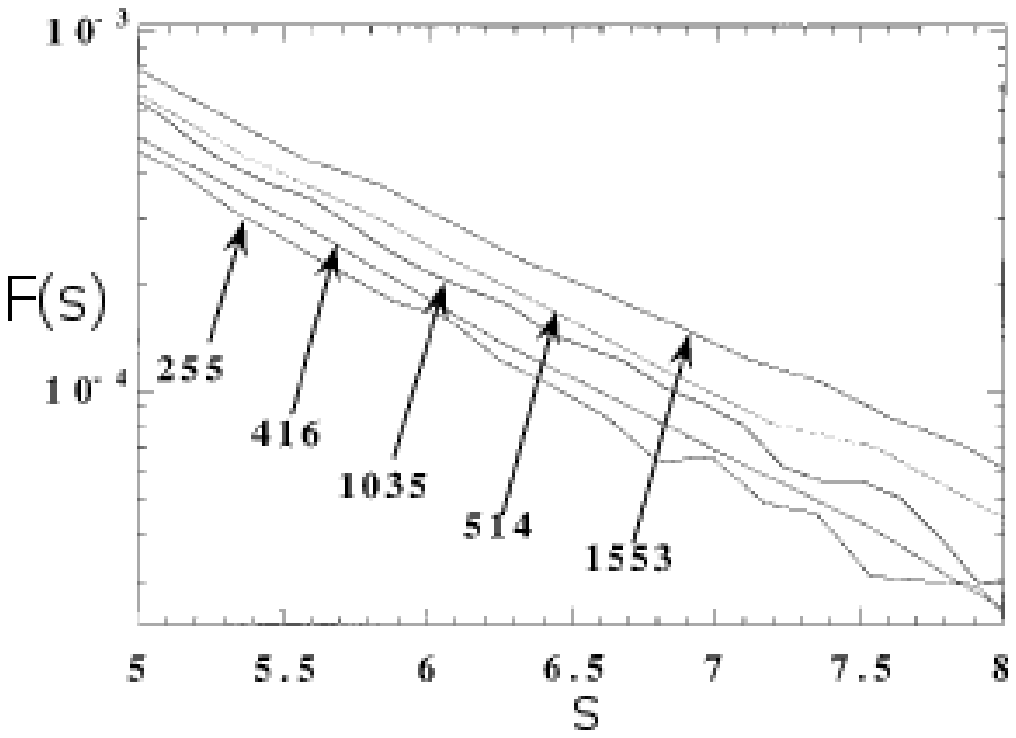}
\vspace{-0.mm}
\caption{PDF of $\partial u_r/\partial r$ for 
$R_{\lambda}$ = 255, 416, 514, 1035 and 1553. These data are from Ref. \cite{Tabeling97} }
\vspace{-0.mm}
\label{tabeling}
\end{figure}
Figure \ref{figura_3} can be compared with the data collected by Sreenivasan and Antonia \cite{Antonia97}, which are here reported into Fig. \ref{antonia}.
These latter are referred to several measurements and simulations obtained in different situations which can be very far from the isotropy and homogeneity conditions. 
Nevertheless a comparison between the present results and those of 
Ref. \cite{Antonia97} is an opportunity to state if the two data exhibit 
elements in common.
According to  Ref. \cite{Antonia97}, the flatness monotonically rises with $R_\lambda$ with a rising rate which agrees with Eq. (\ref{m2}) for 
$10 < R_\lambda < 60$ (dashed line, Fig. \ref{figura_3}), whereas the skewness seems to exhibit minor variations.
Thereafter, $H_4$ continues to rise with about the same rate, without the saturation observed 
in Fig. \ref{figura_3}.
The weaker intermittency calculated with the present analysis seems arise from the isotropy which makes $u_r$  a gaussian random variable, while, as seen in sec. \ref{s6}, without the isotropy, 
the flatness of $u_r$  and $\Delta u_r$  can be much greater than that of the isotropic case.

Next, the obtained results are compared with the data of Tabeling {\it et al} 
\cite{Tabeling96},  \cite{Tabeling97}. There, in an experiment using low temperature helium gas between two counter-rotating cylinders (closed cell), the authors measure the PDF of $\partial u_r/\partial r$ and its moments.
Again, the flow could be quite far from to the isotropy condition, since
these experiments pertain wall-bounded flows, where the walls could importantly influence the fluid velocity in proximity of the probe. 
The authors found that  moments $H_p$, with $p> 3$, first increase with
$R_{\lambda}$ until $R_{\lambda} \approx$ 700, then exhibit a lightly non-monotonic evolution with respect to $R_{\lambda}$, and finally cease their
variations, denoting a transition behavior (See Fig. \ref{tabeling0}). 
As far as the skewness is concerned, the authors observe small percentage variations. 
Although the isotropy does not describe the non-monotonic evolution near 
$R_{\lambda} =$ 700, the results obtained with Eq. (\ref{fluc4}) can be considered comparable with those of Refs. \cite{Tabeling96}, \cite{Tabeling97}, resulting again, 
that the proposed analysis gives a weaker intermittency with respect to 
Refs. \cite{Tabeling96}, \cite{Tabeling97}.

The normalized PDFs of $\partial u_r/\partial r$ are calculated with Eqs. (\ref{frobenious_perron}) and (\ref{fluc4}), and are shown in Fig. \ref{figura_4} 
in terms of the variable $s$, which is defined as  
\bea
\ds s = \frac{\partial u_r/\partial r} 
{ \left\langle (\partial u_r/\partial r)^2\right\rangle^{1/2}  }
\nonumber
\eea
Figure \ref{figura_4}a shows the diagrams for $R_{\lambda} =$ 15, 30 and 60, where
the PDFs vary in such a way that $H_3(0) = -3/7$.
\\
As well as in Ref. \cite{Tabeling97}, Figs. 4b and 4c give the PDF for
$R_{\lambda}$ = 255, 416, 514, 1035 and 1553,  where these last Reynolds numbers are calculated through the  Kolmogorov function given in Ref. \cite{Tabeling97}, with $H_3(0) = -3/7$.
In particular, Fig. \ref{figura_4}c represents the enlarged region of Fig. \ref{figura_4}b,  where the tails of PDF are shown for $5 < s < 8$.
According to Eq. (\ref{fluc4}), the tails of the PDF rise in the interval 
10 $< R_{\lambda} <$ 700,
whereas at higher values of $R_{\lambda}$, smaller variations occur.
Although the trend observed in Fig. \ref{tabeling} \cite{Tabeling97} is non-monotonic, 
Fig. \ref{figura_4}c shows that the values of the PDFs
calculated with the proposed analysis, for $5 < s < 8$, exhibit the same
order of magnitude of those obtained by Tabeling {\it et al} 
\cite{Tabeling97}.
\begin{figure}[t]
\vspace{0.mm}
 \centering
\hspace{-0.mm}
\includegraphics[width=0.50\textwidth]{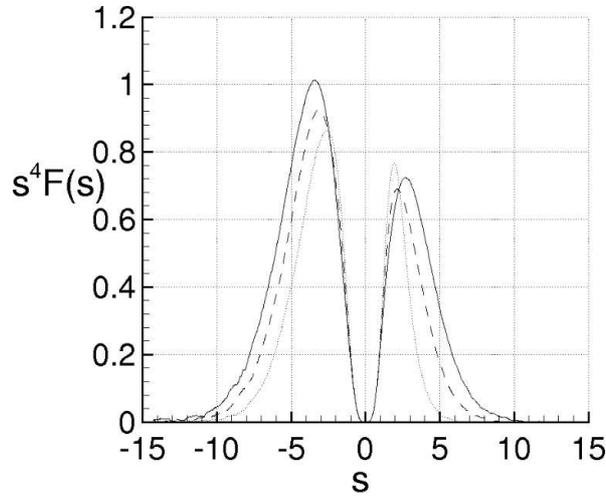}
\vspace{-0.mm}
\caption{Plot of the integrand $s^4 F(s)$ for different
$R_\lambda$. Dotted, dashdotted and continuous lines are
for $R_\lambda$ = 15, 30 and 60, respectively.}
\vspace{-0.mm}
\label{figura_5}
\end{figure}

Asymmetry and intermittency of the distribution functions are also 
represented through the integrand function of the $4^{th}$
order moment of PDF, which is
$
J_4(s) = s^4 F(s)
\nonumber
$.
This function is shown in terms of $s$, in Fig. \ref{figura_5}, for
 $R_\lambda$ = 15, 30 and 60.

\section{\bf  Conclusions \label{s9}}

The proposed analysis is based on the conjecture that  
the turbulence is caused by the bifurcations
and that the kinematic of the relative motion is much more rapid than
the fluid state variables. 
The analysis also assumes statistical hypotheses about the velocity
difference, which derive from the condition of fully developed turbulence. 
The main limitation of this analysis is that it only studies 
the developed homogeneous-isotropic turbulence,
whereas this does not consider the intermediate stages of the turbulence.

The results can be here summarized:

\begin{enumerate}

\item
 The qualitative analysis of the bifurcations leads to determine
the order of magnitude of the critical Reynolds number based on the Taylor scale
and the number of the bifurcations at the onset of the turbulence.

\item
The momentum equations written using the referential coordinates allow
to factorize the velocity fluctuation and to express it in Lyapunov exponential form
of the local fluid deformation. 
As a result, the velocity fluctuation is the combined effect of the exponential growth rate and of the rotations of the Lyapunov basis with respect to the fixed frame of reference.

\item
The finite scale Lyapunov analysis of the relative motion provides
an explanation of the physical mechanism of the energy cascade in turbulence
and gives a closure of the von K\'arm\'an-Howarth equation. 
This is a non-diffusive closure which depends on the local values of
the longitudinal correlation function and on its spatial gradient.

The fluid incompressibility is a sufficient condition to state that the inertia forces transfer the kinetic energy between the length scales without changing the total kinetic energy.
This implies that the skewness of the longitudinal velocity derivative is 
a constant of the present analysis and that the energy cascade mechanism does
not depend on the Reynolds number.

\item
The statistics of $\Delta u_r$ can be inferred looking at the Fourier series of the velocity
difference. This is a non-Gaussian statistics, where the constant skewness of 
$\partial u_r/ \partial r$ implies that the other higher absolute moments 
increase with the Taylor-scale Reynolds number. 

\item
The proposed closure of the von K\'arm\'an-Howarth equation, shows that the mechanism of energy cascade
gives energy spectrum that can be compared with the Kolmogorov law $\kappa^{-5/3}$ in an opportune
range of wave-numbers and which satisfy the incompressibility condition.

\item 
For developed energy spectrum, the Kolmogorov function exhibits, in an opportune 
range of $r$, small variations much less than at the previous times, and its maximum is quite
close to 4/5, whereas the Kolmogorov constant is about equal to 1.93. 
As the consequence, the maximal finite scale Lyapunov exponent and the diffusivity coefficient
vary according to the Richardson law when the separation distance is of the order of the Taylor
scale.

\item
The analysis also determines the scaling exponents of the moments of the
longitudinal velocity difference through a best fitting procedure. 
For developed energy spectrum, these exponents show variations with the moment order
consistent with those present in the literature.
\end{enumerate}

\section{\bf Appendix}

The von K\'arm\'an-Howarth equation gives the evolution in the time of the
longitudinal correlation function for isotropic turbulence.
The correlation function of the velocity components is the symmetrical 
second order tensor
$
\ds R_{i j} ({\bf r}) = \left\langle u_i u_j' \right\rangle 
$, 
where $u_i$ and  $u_j'$ are the velocity components at ${\bf x}$ and 
${\bf x} + {\bf r}$, respectively, being $\bf r$ the separation vector.
The equations for $R_{i j}$ are obtained by the Navier-Stokes equations 
written in the two points ${\bf x}$ and ${\bf x} + {\bf r}$ \cite{Karman38}, \cite{Batchelor53}.
For isotropic turbulence $R_{i j}$ can be expressed as
\bea
R_{i j} ({\bf r}) = u^2 \left[ (f -g) \frac{r_i r_j}{r^2} + g \delta_{i j}\right] 
\label{R_corr}
\eea
$f$ and $g$ are, respectively, longitudinal and lateral correlation functions, which are
\bea
\ds f(r)= \frac{\left\langle u_r({\bf x}) u_r({\bf x}+{\bf r}) \right\rangle }{u^2}, \
\ds g(r)= \frac{\left\langle u_n({\bf r}) u_n({\bf x}+{\bf r}) \right\rangle }{u^2}
\eea
where $u_r$ and $u_n$ are, respectively, the velocity components parallel and normal to  
$\bf r$, whereas $r = \vert {\bf r} \vert$ and 
$u^2$ = $\left\langle u_r^2 \right\rangle$ =$\left\langle u_n^2 \right\rangle$=
$1/3 \left\langle u_i u_i \right\rangle $.
Due to the continuity equation, $f$ and $g$ are linked each other by the relationship
\bea
g = f + \frac{1}{2}  \frac{\partial f}{\partial r} r
\label{g}
\eea

The von K\'arm\'an-Howarth equation reads as follows \cite{Karman38}, \cite{Batchelor53}
\bea
\ds \frac{\partial f}{\partial t}  = 
\ds \frac{K}{u^2}    +
\ds 2 \nu  \left(  \frac{\partial^2 f} {\partial r^2} +
\ds \frac{4}{r} \frac{\partial f}{\partial r}  \right)  
-    10 \nu \frac{\partial^2 f}{\partial r^2}(0) f
\label{vk}  
\eea
where $K$ is an even function of $r$, defined as  \cite{Karman38}, \cite{Batchelor53}
\bea
\left(    r \frac{\partial}{\partial r}  + 3  \right) K(r) =
\ds  \frac{\partial }{\partial r_k} 
\ds  \left\langle u_i  u_i' (u_k - u_k')  \right\rangle 
\label{vk1}  
\eea
and which can also be expressed in terms of the longitudinal triple correlation function
$
\ds k(r)= {\left\langle u_r^2({\bf x}) u_r({\bf x}+{\bf r}) \right\rangle }/{u^3}
$
\bea
K(r)= u^3 \left(  \frac{\partial}{\partial r}  + \frac{4}{r}  \right) k(r)
\label{kk}
\eea

The boundary conditions of Eq. (\ref{vk}) are \cite{Karman38}, \cite{Batchelor53}
\bea
f(0) = 1, \ \ \lim_{r \rightarrow \infty} f(r) = 0
\label{bc}
\eea
The viscosity is responsible for the decay of the turbulent kinetic energy, the rate of which is
\cite{Karman38}, \cite{Batchelor53}
\bea
\frac{d u^2} {d t} = 10 \nu u^2 \frac{\partial^2 f}{\partial r^2}(0) 
\label{ke}
\eea
This energy is distributed at different wave-lengths according to the energy spectrum 
$E(\kappa)$ which is calculated as the Fourier Transform of $f u^2$, whereas 
the ''transfer function'' $T(\kappa)$ is the Fourier Transform of $K$ \cite{Batchelor53}, i.e.
\bea
\hspace{-1.0mm}
\left[\begin{array}{c}
\hspace{-1.0mm} \ds E(\kappa) \\\\
\hspace{-1.0mm} \ds T(\kappa)
\end{array}\right]  
\hspace{-1.5mm}= 
\hspace{-1.5mm} \frac{1}{\pi} 
\hspace{-1.0mm} \int_0^{\infty} 
\hspace{-1.5mm}\left[\begin{array}{c}
\hspace{-1.0mm} \ds  u^2 f(r) \\\\
\hspace{-1.0mm} \ds K(r)
\end{array}\right]  \kappa^2 r^2 \hspace{-1.0mm} 
\left( \hspace{-0.5mm}\frac{\sin \kappa r }{\kappa r} - \cos \kappa r \hspace{-0.5mm} \right) d r 
\label{Ek}
\eea
where $\kappa = \vert {\bf \bfkappa} \vert$ and $T(\kappa)$ identically 
satisfies to the integral condition
\bea
\int_0^\infty T(\kappa) d \kappa = 0
\label{tk0}
\eea 
which states that $K$  does not modify the total kinetic energy.
The rate of energy dissipation $\varepsilon$ is calculated for 
isotropic turbulence as follows \cite{Batchelor53}
\bea
\ds \varepsilon = -\frac{3}{2} \frac{d u^2} {d t}=
  2 \nu \int_0^{\infty} \kappa^2 E(\kappa) d \kappa
\eea
The microscales of Taylor $\lambda_T$, and of Kolmogorov $\ell$, are defined as
\bea
\begin{array}{c@{\hspace{+0.2cm}}l}
\ds \lambda_T^2 = \frac{u^2}{\langle (\partial u_r/\partial r)^2 \rangle} = 
-\frac{1}{\partial^2 f/\partial r^2(0)}, \ 
\ds \ell = \left( \frac{\nu^3} { \varepsilon}\right)^{1/4}
\end{array}
\eea

\bigskip



\end{document}